\documentclass[aps,prb, notitlepage, singlecolumn, preprintnumbers, nofootinbib, 
nolongbibliography,
amsmath,amssymb,superscriptaddress,floatfix,scrartcl]{revtex4-2}

\usepackage{amssymb}
\usepackage{color}
\usepackage{graphicx}
\usepackage{mathrsfs}
\usepackage[unicode=true,pdfusetitle,bookmarks=false,colorlinks=true,citecolor=blue,urlcolor=blue,linkcolor=red]{hyperref}
\usepackage{bm}
\usepackage{braket}

\makeatletter
\def\widebar{\accentset{{\cc@style\underline{\mskip10mu}}}}
\makeatother
\makeatletter
\def\wideubar{\underaccent{{\cc@style\underline{\mskip10mu}}}}
\makeatother

\makeatletter

\@addtoreset{equation}{section}
\makeatother

\newcommand{\normord}[1]{%
  \ddagger\!\mathrel{\mspace{2mu}#1\mspace{2mu}}\!\ddagger%
}

\newcommand{\ii}{\mathrm{i}}
\newcommand{\dd}{\mathrm{d}}

\newcommand{\bx}{{\bm{x}}}
\newcommand{\bk}{{\bm{k}}}
\newcommand{\Z}{{\mathbb{Z}}}

\usepackage[normalem]{ulem} 

\begin{document}

\preprint{YITP-23-105}


\title{
Higher bracket structure of
density operators
in Weyl fermion systems and 
topological insulators}

\author{Edwin Langmann}
\email{langmann@kth.se}
\affiliation{Department of Theoretical Physics, KTH Royal Institute of Technology, SE-106 91 Stockholm, Sweden}
\author{Shinsei Ryu}
\email{shinseir@princeton.edu}
\affiliation{Department of Physics, Princeton University, Princeton, New Jersey 08544, USA}
\author{Ken Shiozaki}
\email{ken.shiozaki@yukawa.kyoto-u.ac.jp}
\affiliation{Center for Gravitational Physics and Quantum Information, Yukawa Institute for Theoretical Physics, Kyoto University, Kyoto 606-8502, Japan}

\date{\today}

\begin{abstract}
We study the algebraic structure of electron density operators in 
gapless Weyl fermion systems in $d=3,5,7,\cdots$ spatial dimensions 
and in topological insulators (without any protecting symmetry) 
in $d=4,6,8,\cdots$ spatial dimensions.
These systems are closely related by the celebrated bulk-boundary correspondence.
Specifically, we study the higher bracket 
-- a generalization of commutator for more than two operators --
of electron density operators in these systems. 
For topological insulators, 
we show that 
the higher-bracket algebraic structure of density operators 
structurally parallels with the
Girvin-MacDonald-Platzman algebra (the $W_{1+\infty}$ algebra), 
the algebra of electron density operators projected onto the lowest Landau level in the quantum Hall effect.
By the bulk-boundary correspondence,
the bulk higher-bracket structure mirrors its counterparts at the boundary.
Specifically, 
we show that 
the density operators of Weyl fermion systems, 
once normal-ordered with respect to the ground state, 
their higher bracket acquires a c-number part. 
This part is an analog of the Schwinger term in the commutator of the fermion current operators. We further identify this part with a cyclic cocycle, which is a topological invariant and an element of 
Connes' noncommutative geometry.
\end{abstract}

\pacs{??} 

\maketitle

\tableofcontents

\section{Introduction}

Topological insulators represent a 
class of condensed matter systems that defy conventional characterization. 
While fully gapped in the bulk, they support gapless excitations 
along their surface or edge. 
The key to their remarkable behavior lies in the underlying topology of the quantum states, where topological invariants dictate the emergence of robust, protected surface states immune to scattering and disorder. 
Examples of topological insulators include, in a broad context,
the quantum Hall effect (QHE) in two spatial dimensions,
and its lattice incarnation, Chern insulators,
the quantum spin Hall effect,
time-reversal symmetric topological insulators in three spatial dimensions,
and so forth.
In the last decade, topological insulators have been a central topic
in condensed matter physics, 
and many examples have been discovered experimentally
\cite{HasanKane2010, QiZhang2011, Chiu_2016}.

In this paper, we consider the algebra of 
electron density operators
in topological band insulators in their bulk and also at their boundaries.
At the single-particle level,
previous works explored 
the higher-bracket structure 
of the projected electron position operators
in bulk topological insulators
\cite{Neupert_2012,Estienne_2012,Shiozaki_2013,
Hasebe_2014b,Hasebe_2014a,Hasebe_2017,
hasebe2023generating} 
-- see below for more descriptions of the higher bracket structure.
The scope of this paper is 
its many-body incarnations,
which may open a door 
into a collective description 
of topological insulators
that may be able to
incorporate interaction effects
and fractional counterparts of topological insulators.

\subsection{Review of the density operator algebra in the QHE}
The motivation and the scope of the paper
is best described by briefly reviewing the physics of the (2+1)d QHE.

It is well known that one of the most fundamental properties of the QHE is the non-commutativity of electron coordinates:
Once the dynamics of electrons is constrained within a given (e.g., the lowest) 
Landau level, their position operators do not commute,
\begin{align}
[{X},Y]  =-\ii \ell^2_0,
\label{XY}    
\end{align}
where $X$ and $Y$ are the $x$ and $y$ components of the single-particle
(first quantized)
electron coordinate operator projected onto the lowest Landau level (LLL),
and $\ell_0$ is the magnetic length. 
One can also consider arbitrary functions
of electron coordinates $f(x,y)$
and their projected counterparts, $f({X}, Y)$.
Once properly (anti-normal) ordered,
the commutator of two such operators
$f(X, Y)$ and $g(X, Y)$ satisfies 
\begin{align}
\left[f({X},{Y}),g({X},{Y})\right]
= \{\!\{f,g\}\!\}(X,Y), 
\label{moyal}
\end{align}
where
$\{\!\{\cdots \}\!\}$
is the Moyal bracket defined by 
\begin{align}
\{\!\{f, g\}\!\}
:=
 - 2\sum_{n=0}^\infty \frac{(\ii \ell_0^2/2)^{2n+1} }{(2n+1)!}\left( (\partial_x^{2n+1}f)(\partial_y^{2n+1}g) - (\partial_y^{2n+1}f)(\partial_x^{2n+1}g)  \right)  .
  \nonumber 
\end{align}
Note that to the lowest order in $\ell_0$, the Moyal bracket simplifies to the Poisson bracket, 
\begin{align}   \label{PB}      
\left[  f({X},{Y}), g({X}, {Y})\right]
=      -\ii \ell^2_0 \{ f, g\}_{\mathrm{PB}}(X,Y) +\cdots, 
\end{align}
where
$\{ f, g\}_{\mathrm{PB}} = (\partial_x f)(\partial_y g)-(\partial_y f)(\partial_x g)$.

The coordinate non-commutativity
is translated into a non-trivial algebra obeyed by the electron density operators:
The function $f(x,y)$ defines a
corresponding many-body ({\it second quantized}) operator through
\begin{align}
\hat{\rho}(f) := \int_{M_2} \dd^2x\, f(x,y)
\hat{\rho}(x,y),
\end{align}
where $\hat{\rho}(x,y)$ is the electron density operator projected to the lowest Landau level,
and the integral is over the 2d spatial manifold $M_2$ that hosts the droplet of the electron liquid.
$\hat{\rho}(f)$ satisfies what is known as the Girvin-MacDonald-Platzman (GMP) algebra, 
\begin{align}
\left[\hat{\rho}(f_1), \hat{\rho}(f_2)\right] 
=
\hat{\rho}(\{\!\{f_1, f_2\}\!\}).
  \label{bulk alg 2d}
\end{align}
This density operator algebra (or its counterpart in terms of electron coordinates) is also known as
the $W_{1+\infty}$ algebra or the Fairlie-Fletcher-Zachos algebra
\cite{Girvin_1985,Girvin_1986, 
FAIRLIE1989203, FAIRLIE1989101, FLETCHER1990323, Iso_1992, Cappelli_1993, Martinez_Stone_1993}.
The GMP algebra plays a crucial role in studying,
e.g., bulk charge neutral collective excitations in the fractional quantum Hall effect.
(Our focus in this paper however will be edge excitations and integer filling.)
By taking
$f(x,y) = e^{\ii \boldsymbol{q} \cdot \boldsymbol{r}}$,
Eq.\ \eqref{bulk alg 2d}
reduces to an alternative, but equivalent form of the GMP algebra,
written in terms of the Fourier modes
of the projected density operator $\hat{\rho}(\boldsymbol{q})$:
\begin{align}
[\hat{\rho}(\boldsymbol{q}_1),\hat{\rho}(\boldsymbol{q}_2)]
&= 2 \ii \sin \left(
(\ell^2_0/2)
\boldsymbol{q}_1\wedge \boldsymbol{q}_2
\right) 
\hat{\rho}(\boldsymbol{q}_1+\boldsymbol{q}_2)
\label{GMP0} 
\end{align}
where
$\boldsymbol{q}\wedge \boldsymbol{q}'
=
q^{\ }_xq_y'-q^{\ }_yq_x'
$. 
\footnote{
  In the literature, one often finds an extra factor of
  $e^{(\ell^2_0/2) \boldsymbol{q}_1 \cdot \boldsymbol{q}_2}$
  in the RHS of Eq.\ \eqref{GMP0}.
  This factor arises as a multiplicative
  renormalization when one anti-normal orders
  the single-particle operator $e^{\ii \boldsymbol{q}\cdot \boldsymbol{r}}$.
  See also the discussion around Eq.\ \eqref{matrix elem density op}
  in the language of second quantization.
}

Let us now turn our attention to the boundary of the quantum Hall droplet.
The density operators at the boundary of the droplet 
obey the $U(1)$ current (Kac-Moody) algebra
\begin{align}
  \displaystyle
  [ \hat{\varrho}(f_1), \hat{\varrho}(f_2)]
  =
  \frac{-\ii \nu}{2\pi} \int_{\partial M_2} f_1 \dd{f}_2,
      \label{bdry algebra}
\end{align}
where $\nu$ is the filling fraction,
$\hat{\varrho}(f)$ is the boundary electron density operator
weighted by an envelope function $f(x)$,
$\hat{\varrho}(f)= \int_{\partial M_2} \dd{x}\, f(x)\hat{\varrho}(x)$,
where $x$ parameterizes the flat one-dimensional space $\partial M_2$, the boundary of the bulk manifold $M_2$.
\footnote{
We use $\hat{\rho}$ and $\hat{\varrho}$
to denote the bulk and boundary density operators, respectively.
The bulk density operator is 
un-normal-ordered, and obtained 
by projecting the electron density operator to a given set of (topological) bands.
The boundary density operator 
is normal-ordered with respect to the ground state of the gapless 
boundary Hamiltonian.
We will use $\boldsymbol{r}$ and $\boldsymbol{q}$ etc. to denote 
the $d$-dimensional bulk spatial coordinates
and momentum. 
On the other hand, 
we will use ${\bf r}$ and ${\bf q}$ etc. to denote 
the $(d-1)$-dimensional 
boundary spatial coordinates
and momentum.}
Taking $f_{\alpha=1,2}(x)=\delta(x-x_\alpha)$, the $U(1)$ current algebra can also be written as
\begin{align}
  [\hat{\varrho}(x_1), \hat{\varrho}(x_2)]
    &=
      \frac{-\ii \nu}{2\pi} 
      \partial_{x_1} \delta(x_1-x_2),
      \label{bdry algebra 1}
\end{align}
where $x_{1,2}$ is the coordinate along the edge.
The right-hand side is known as the Schwinger term.
The density operator algebra Eq.\ \eqref{bdry algebra}
can be derived from the non-interacting chiral edge mode.
For integer filling, $\nu=1$, say, it is described by the Hamiltonian
\begin{align}
  \label{edge Ham}
  \hat{H} = \int_{\partial M_2} \dd{x}\, \normord{\hat{\psi}^{\dag} (-\ii \partial_x )\hat{\psi}},
\end{align}
where $\hat{\psi}/\hat{\psi}^{\dag}$ is the fermion annihilation/creation operator
for Weyl fermion modes.
Here, $\normord{\cdots}$ represents normal ordering
with respect to the ground state of Eq.\ \eqref{edge Ham}.
The fermion density operator 
$\hat{\varrho}(x) = \normord{\hat{\psi}^{\dag} \hat{\psi}}(x)$,
once properly normal-ordered with respect to the ground state, 
obeys the $U(1)$ current algebra.
       
The bulk and boundary density operator algebras are closely related. 
To derive the boundary algebra \eqref{bdry algebra}
from the bulk algebra \eqref{bulk alg 2d},
we follow Ref.\ \cite{Martinez_Stone_1993},
and set
$f_{\alpha=1,2}(x,y)= f_\alpha(x) g(y)$ in Eq.\ \eqref{bulk alg 2d}:
\begin{align}
&
\left[
{\hat{\rho}}(f_1), {\hat{\rho}}(f_2)
\right] 
\nonumber \\
&\quad =
\frac{\ii  \ell^2_0 }{2}
\int \dd{x}
\int^{+\infty}_{-\infty} \dd{y}\, 
\left[
(\partial_x f_1) f_2
-
(\partial_x f_2) f_1
\right] 
\partial_y g^2
\hat{\rho}(x,y). 
\end{align}
To capture dynamics at the edge, we take an envelope function $g(y)$
such that $g(y)\to 0$ at $y\to \pm \infty$
and 
$g(y)=1$ around $y=0$,
resulting in
$\left[
  {\hat{\rho}}(f_1), {\hat{\rho}}(f_2)
\right] 
\sim -\ii\rho_0 \ell^2_0 \int_{\partial M_2} f_1 \dd{f}_2$.
Clearly, the result does not depend on the details of this function as long as
$\partial_y g \to 0$ as $y\to \pm\infty$.
Furthermore, we replace the density operator
by its expectation value inside the droplet,
$\hat{\rho}
\to
\langle \hat{\rho}
\rangle=:\rho_0$ for $y<0$,
and note
$\rho_0 = \nu/\ell_0^2$.
After renaming $\hat{\rho}\to \hat{\varrho}$,
we recover the boundary algebra \eqref{bdry algebra}.

There is an alternative derivation of
the boundary current algebra from the bulk density operator algebra, 
in which the effect of the boundary
to the many-body ground state 
is more explicitly taken into account
\cite{Azuma_1994, Cappelli_2018}.
In the presence of a (smooth) confining potential, the Landau levels are (weakly) dispersed, and
a finite droplet of an integer quantum Hall state is formed
by filling all Landau levels below the chemical potential. 
Once the ground state is specified,
operators have to be {\it normal ordered} with respect to the ground state: 
For any quadratic operator of the form 
$
  \sum A_{nm}\hat{c}^{\dag}_n \hat{c}^{\ }_m
  $
where $\hat{c}^{\dag}_n/\hat{c}^{\ }_n$ are the fermion creation/annihilation operators for states within the LLL,
we consider its normal-ordered counterpart
$
  \sum A_{nm}
  \normord{\hat{c}^{\dag}_n \hat{c}^{\ }_m} $.
Once normal-ordered density operators are considered,
the GMP algebra acquires a central extension term.
Expanded in small magnetic length,
this is nothing but the boundary current algebra
--
see Section \ref{The Wannier functions and the bulk-boundary relation}
for more details.

We close the brief review of the (2+1)d QHE with some comments.

First, the GMP algebra and the bulk-boundary correspondence of the QHE
can be generalized to Chern insulators or anomalous quantum Hall systems, 
i.e.,  
(2+1)d lattice systems exhibiting the QHE even in the absence of a uniform applied magnetic field.
In Chern insulators, 
the GMP algebra does not arise in its original form due to the non-uniform distribution of the Berry curvature
in momentum space.
Nevertheless, once quantum Hall liquid is formed, it has been observed that the GMP algebra
emerges at low energies, 
and it is used to identify the topological nature of (fractional) Chern insulators
\cite{Parameswaran2012,Parameswaran2013,Bernevig2012}.

Second, the $U(1)$ current algebra
\eqref{bdry algebra}
is an example of affine Lie algebras (Kac-Moody algebra),
which are 
infinite-dimensional extensions 
of finite-dimensional Lie algebras.
What appears on the RHS of Eq.\ \eqref{bdry algebra} is the Schwinger term
that arises due to regularization 
to remove ultraviolet divergences.
The Schwinger term is an example of the Kac-Peterson cocycle,
which is a specific 2-cocycle 
and plays a crucial role in the classification of the central extensions of Kac-Moody algebras. 
The current algebras, both Abelian and non-Abelian ones,  
appear in various contexts including  
(1+1)d many-body quantum systems at criticality 
and play a vital role as spectrum generating algebra.
In this paper, our primary focus remains on the Abelian case.  
Nevertheless, our results apply also to non-Abelian cases 
(the interested reader can find details in Section \ref{sec:nonAbelian}).

\subsection{Higher-dimensional generalization}

Guided by the above algebraic structure in the QHE, 
in this paper, we will seek an analogous algebra of the electron density operators 
in higher dimensional topological insulators.
In higher-dimensional topological insulators,
there is growing evidence that non-commutative geometry may also play an important role
\cite{Neupert_2012,Estienne_2012,Shiozaki_2013,
Hasebe_2014b,Hasebe_2014a,Hasebe_2017,
hasebe2023generating}. 
The relevant non-commutativity, however, involves an $n$-ary structure
or $n$-bracket structure with $n>2$.
For example, in (4+1)d class A topological insulators, the relevant
coordinate algebra involves the four-bracket of the projected electron coordinates,
\begin{align}
[X_{1}, X_{2}, X_{3}, X_{4}]
:=
\epsilon^{\alpha\beta\gamma\delta}
  X_{\alpha} X_{\beta} 
  X_{\gamma} X_{\delta},
  \label{4-bracket in qm}
\end{align}
where $X_{\alpha}$ ($\alpha=1,2,3,4$) are the $\alpha$-th components
of the (single-particle or first quantized) electron coordinate operator projected onto
the occupied Bloch bands of topological insulators.
The four-bracket structure can be motivated by
recalling that, in momentum space, the projected position operators are covariant derivative 
$X_{\alpha}= \ii \partial/\partial k_{\alpha}- {\cal A}_{\alpha}(\boldsymbol{k})$
where ${\cal A}_{\alpha}(\boldsymbol{k})$ is the (non-Abelian) Berry connection
in momentum space.
Then, the four-bracket is given by
the topological density  
$\epsilon^{\alpha\beta\gamma\delta}{\cal F}_{\alpha\beta}(\boldsymbol{k})
{\cal F}_{\gamma\delta}(\boldsymbol{k})$,
where ${\cal F}_{\alpha\beta}(\boldsymbol{k})=[X_{\alpha}, X_{\beta}]$ is the Berry curvature.
As an example, for the quantum Hall state in (4+1)d, the four-bracket is proportional to the identity operator and given by
\cite{Hasebe_2014a}
$[X_{1}, X_{2}, X_{3}, X_{4}]
  =
  -\ell^4_0$,
in complete analogy to the (2+1)d case. 
(This algebra is obtained for the (4+1)d quantum Hall state  
realized on the 4d sphere, and taking the flat space limit around the north pole.
Here, the ``magnetic length'' $\ell_0$ is roughly related to the radius of the 4d sphere normalized by the strength of the
monopole placed ``inside'' of the sphere.)
The algebraic structure
of the type \eqref{4-bracket in qm}
is known as higher-order Lie algebra; see, for example,
Ref.\ \cite{de_Azc_rraga_1997}.

Following the discussion of the (2+1)d QHE above,
for functions of the coordinates $X_{\alpha}$,
$f_{\alpha=1,2,3,4}(X_1, X_2, X_3, X_4)$, 
we expect that
their four bracket
$[f_1, f_2, f_3, f_4]$ reflects the non-trivial 
four-bracket structure of the coordinates.
We note that to properly define $f_\alpha$ as a quantum mechanical operator,
we need to introduce a proper ordering of operators, which is
not clear at this moment. Nevertheless, to the lowest order in $\ell_0$,
we expect 
\begin{align}
  \label{4d density single}
  [f_1, f_2, f_3, f_4]
  &\sim
  \{f_1, f_2, f_3, f_4\}
  \nonumber \\
  &=
  \epsilon^{\alpha\beta\gamma\delta}
  \partial_{\alpha}f_1
  \partial_{\beta}f_2
  \partial_{\gamma}f_3
  \partial_{\delta}f_4,
\end{align}
where what appears on the RHS,
$\{f_1, f_2, f_3, f_4\}$,
is the so-called Nambu bracket,
a generalization of the Poisson bracket
\cite{Nambu1973}.
Unlike the case of the Moyal product in (2+1)d, the structures at higher orders are unclear. 

The $n$-bracket algebra of the projected electron coordinate operators
at the level of single-particle quantum mechanics
is an interesting higher structure.
Following once again the analogy with the (2+1)d case,
we may expect that the $n$-bracket algebra also plays an important role 
at the level of many-body quantum physics.  
In this paper, we consider a fully antisymmetrized product of four projected fermion density operators,
both in the bulk and boundary of higher-dimensional 
topological insulators.
For presentational simplicity, we mostly focus on the case of four spatial
dimensions here.  
Specifically, we introduce the four-bracket for four second-quantized operators acting
on the fermion Fock space (many-body Hilbert space).
For example, for the bulk (projected) density operator 
$\hat{\rho}(f_\alpha)$, we consider 
\begin{align}
&
  \big[\hat{\rho}(f_1),
  \hat{\rho}(f_2),
  \hat{\rho}(f_3),
  \hat{\rho}(f_4)\big]_{mod}
  \nonumber \\
 &\quad :=
\epsilon^{\alpha\beta\gamma\delta}
\hat{\rho}(f_\alpha)
\hat{\rho}(f_\beta)
\hat{\rho}(f_\gamma)
\hat{\rho}(f_\delta)
-
(\mbox{two-body terms}).
\label{rough def mod bracket}
\end{align}
Here, the first term is the  
direct analogue of Eq.\ \eqref{4-bracket in qm} defined with the single-particle
Hilbert space.
In the second term, we subtract some two-body terms:
The precise definition of the subtraction term can be found
in Section \ref{n-bracket and second quantization}.
We call this object the modified 4-bracket.
Following the discussion in the QHE,
the above algebra \eqref{4d density single} implies that 
$\big[\hat{\rho}(f_1),
\hat{\rho}(f_2),
\hat{\rho}(f_3),
\hat{\rho}(f_4)\big]_{{mod}}
\sim \hat{\rho}\big(\{f_1,f_2,f_3,f_4\}\big)$.
Setting $f_\alpha(x_1,x_2,x_3, x_4)= f_\alpha(x_1,x_2,x_3) g(x_4)$ 
and assuming the same profile of $g(x_4)$ as in the case of the QHE,
and the constant density in the bulk, we would then conclude the density operator algebra
on the surface of a (4+1)d topological insulator is 
\begin{align}
&
\left[
\hat{\rho}(f_1), \hat{\rho}(f_2),
\hat{\rho}(f_2), \hat{\rho}(f_3)
\right] 
\sim 
\int_{\partial M_4}
f_1 \dd{f}_2 \dd{f}_3 \dd{f}_4.  
\end{align}
This argument suggests the existence of an analog of the Schwinger term
if one considers the four-bracket of the electron density operator
on the surface of a topological insulator. 
Guided by this observation, one of our main goals in this paper is to examine this idea
by developing calculations similar to
the purely boundary calculation of the Schwinger term in the (2+1)d QHE.

\subsection{Organization and summary of the paper}
Let us now outline our calculations and main results.  
We start in Section \ref{sec:2ndquantizationmap}
by recalling some known facts about second quantization and normal ordering.
The formalism reviewed there
(as well as its higher-bracket counterparts developed later in
Sections \ref{sec:nbrackets} and \ref{sec:higherdim})
is used throughout the paper,
both for the bulk and boundary density operator algebras.
For the rest of Section \ref{n-bracket and second quantization}, we focus on the boundaries of topological insulators. 
Specifically, we consider the density operator algebra of
the Weyl fermion theory in even spacetime dimensions,  
such as the (1+1)d edge theory \eqref{edge Ham}
or the (3+1)-dimensional Weyl fermion theory
with the Hamiltonian,  
\begin{align}
  \hat{H}_{{\it Weyl}} =\int_{\partial M_4} \dd^3\mathbf{x}\, \sum_{i=1}^3
  \normord{\hat{\psi}^{\dag} (-\ii \sigma_{i}\partial_{i} ) \hat{\psi}},
  \label{3+1 Weyl intro}
\end{align}
where $\hat{\psi}^{\dag},\hat{\psi}$ are the two-component creation/annihilation operators
of the Weyl fermion mode, and 
$\sigma_{i=1,2,3}$ are the $2\times 2$ Pauli matrices.
These theories appear on the boundaries of (2+1)- and (4+1)-dimensional
class A topological insulators, respectively. 

In Section \ref{sec:NCG}, using the (1+1)-dimensional edge theory \eqref{edge Ham} as an example,  we shortly describe a mathematical interpretation of the Schwinger term in 
Eqs.\ \eqref{bdry algebra} and \eqref{bdry algebra 1}
in the context of noncommutative geometry (NCG)
\cite{Connes1994,Gracia2001}.
This short intercourse guides us regarding how to generalize the Schwinger term and the boundary density operator algebra
in higher dimensions,
as described in Sections \ref{sec:nbrackets} and \ref{sec:higherdim}.

Taking the (3+1)-dimensional Weyl fermion theory \eqref{3+1 Weyl intro}
as an example, let us describe the main result 
of Section~\ref{sec:nbrackets}.
As in the case of (1+1) dimensions, we consider the normal-ordered density operator
$\hat{\varrho}({\bf x}) = \normord{\hat{\psi}^{\dag} \hat{\psi}}({\bf x})$
of the boundary system \eqref{3+1 Weyl intro},
and also the smeared density operator
$\hat{\varrho}(f)= \int \dd^3{\bf x}\, f({\bf x}) \hat{\varrho}({\bf x})$.
The 4-bracket of the density operators is given by
\begin{align}
  &
    [\hat{\varrho}(f_1), \hat{\varrho}(f_2) ,\hat{\varrho}(f_3), \hat{\varrho}(f_4)]
    \nonumber \\
  &\quad =
    \epsilon^{\alpha\beta\gamma\delta}
    \hat{\varrho}(f_{\alpha}) \hat{\varrho}(f_{\beta}) \hat{\varrho}(f_{\gamma}) \hat{\varrho}(f_{\delta})
    \nonumber \\
  &\quad = \hat{\varrho}([f_1,f_2,f_3,f_4])
      + b_4([f_1,f_2,f_3,f_4])
      + \epsilon^{\alpha\beta\gamma\delta} 
      \hat{Q}_{0,2}(
      f_{\alpha}f_{\beta}\otimes f_{\gamma}f_{\delta}),
\end{align}
Here, $\hat{Q}_{0,2}$ is some four-fermion operator.
The part which is of most interest is
$b_4$, which is a 
linear c-number valued function of the four-bracket coming from normal ordering.
As we will see, this part gives rise to a proper
generalization of the Schwinger term.
In the analogous (same) notation, the $U(1)$ current algebra
\eqref{bdry algebra}
in (1+1)d 
arises as follows,
$[\hat{\varrho}(f_1), \hat{\varrho}(f_2)]
  = \hat{\varrho}([f_1,f_2])
    +
    b_2([f_1,f_2])
$, 
where the $c$-number part  $b_2$ comes from normal ordering; 
the non-trivial (well-known) result is that, 
when computed carefully (with proper regularizations), this gives rise to a finite nontrivial Schwinger term.

As explained in Section \ref{sec:nbrackets},
the connection with noncommutative geometry motivates us to split $b_4$ into two parts,
\begin{align}
b_4([f_1,f_2,f_3,f_4]) &= S_4(f_1,f_2,f_3,f_4) + R_4(f_1,f_2,f_3,f_4).
\end{align}
The rationale for this splitting is that 
$S_4$ is given in terms of 
traces of trace-class operators
(in the single particle Hilbert space,
i.e.,
the Hilbert space of square-integrable functions on 3d space $M_3$
with spin 1/2 degrees of freedom),
while $R_4$ is not.
Hence, $S_4$ is robust (insensitive)
to the choice of the UV cutoff,
while $R_4$ is not.
The first term $S_4$ was evaluated in Ref.\ \cite{Langmann1995} and is given by
\begin{align}
  \displaystyle
  S_4(f_1,f_2,f_3,f_4)
  =
  \frac{1}{2 \pi^2} \int_{\partial M_4} 
  f_1 \dd{f}_2 \dd{f}_3 \dd{f}_4.
  \label{S40 2}
\end{align}
This can be thought of as a proper generalization of the Schwinger term in (1+1)d,
the RHS of Eq.\ \eqref{bdry algebra 1}.
Given 
the fragility of $R_4$, 
i.e., sensitivity to the choice of the UV cutoff, 
we are tempted 
to remove (subtract) $R_4$ and 
interpret this as 
a further renormalization of the 4-bracket of densities 
(in addition to removing 4-body terms).
This proposal is motivated, if not dictated, by the following general principle in quantum field theory:
only quantities that are independent of regularization details are of physics interest. 
Importantly,
as we show in 
Section \ref{sec:nbrackets},
this renormalization does 
not spoil the consistency of the algebra of densities 
with the (renormalized) 4-bracket.
The resulting (3+1)d density operator algebra is 
\begin{align} 
[\hat{\varrho}(f_1),\hat{\varrho}(f_2),\hat{\varrho}(f_3),\hat{\varrho}(f_4)]_{mod} & = \frac{1}{2\pi^2}\int_{\partial M_4} f_1 \dd{f}_2 \dd{f}_3 \dd{f}_4. 
\end{align}

While we explain our construction in detail for the (3+1)d case, 
it is straightforward to generalize our results to arbitrary even spacetime dimensions. 
In particular, the $d$-dimensional Abelian current algebra is given by 
\begin{align} 
\label{S40 3} 
[\hat{\varrho}(f_1), \cdots,\hat{\varrho}(f_{d})]_{mod} = 
-\left(
d/2
\right)!
\left(\frac{\ii}{2\pi}\right)^{d/2}  \int_{\partial M_{d}} f_{1}\dd{f}_{2}\cdots \dd{f}_{d}. 
\end{align} 
See Section \ref{sec:higherdim} for details. 

In Section \ref{Bulk density operator algebra}, we discuss the density operator algebra in the bulk 
of topological insulators in generic $d=2n={\it even}$ dimensions. 
There, we consider the electron density operator 
projected to a given set of (topological) bands.  
The final result for $d={\it even}$ dimensional topological insulators is 
\begin{align}
\label{4d density a}
&
\big[ \hat{\rho}(\boldsymbol{q}_1),
\hat{\rho}(\boldsymbol{q}_2), 
\cdots,
\hat{\rho}(\boldsymbol{q}_d)
\big]_{{mod}}
\nonumber \\
&
\quad =
(\boldsymbol{q}_1 \wedge \cdots \wedge \boldsymbol{q}_d)
\sum_{\hat a, \hat b=1}^{N_-} \Omega
\int_{\mathrm{BZ}} \frac{\dd^{d}\boldsymbol{k}}{(2\pi)^{d}}
\epsilon^{\mu_1 \dots \mu_d}
\nonumber \\
&\quad
\quad \times
\sum_{\hat c_1, \dots, \hat c_{d/2-1} = 1}^{N_-}
  \frac{\mathcal{F}^{\hat a \hat c_1}_{\mu_1\mu_2}}{2} \cdots
  \frac{\mathcal{F}^{\hat c_{d/2-1} \hat b}_{\mu_{d-1}\mu_d}}{2}
\hat{\chi}^{\dag}_{\hat a}(\boldsymbol{k}) \hat{\chi}_{\hat b}(\boldsymbol{k}+\boldsymbol{q}_1+\cdots +\boldsymbol{q}_d) + O(q^{d+1}).
\end{align}
Here, $\big[ \cdots \big]_{{mod}}$ is the modified $d$-bracket mentioned in
Eq.\ \eqref{rough def mod bracket} 
\footnote{
Recall that
the bulk density operator is not normal-ordered
and there is no c-number part 
in Eq.\ \eqref{4d density a},
unlike the boundary density operator algebra
\eqref{S40 3}. 
Hence, we do not need the additional 
renormalization mentioned above.},
$\hat{\chi}^{\dag}$ and $\hat{\chi}$ represent 
electron creation and annihilation operators of the topological bands of interest,
$\Omega$ is the volume of the unit cell, 
hatted indices run over occupied bands
and 
$
{\cal F}^{\hat{a}\hat{b}}
=
(1/2) {\cal F}^{\hat{a}\hat{b}}_{\mu_1 \mu_2}
dk^{\mu_1} dk^{\mu_2}
$
is the Berry curvature -- 
see Section \ref{Bulk density operator algebra} for notations and details.
Provided the Berry curvature is constant in momentum space,
Eq.\ \eqref{4d density a} reduces to
\begin{align}
  \big[ \hat{\rho}(\boldsymbol{q}_1),
                  \cdots,
                  \hat{\rho}(\boldsymbol{q}_d)
  \big]_{{mod}}
  &=
   \rho^{-1}_0 
  (\boldsymbol{q}_1 \wedge \cdots \wedge \boldsymbol{q}_d)
  (d/2)!
    \left(\frac{-\ii}{2\pi} \right)^{{d}/{2}}
    {\it Ch}_{{d}/{2}}\,
    \hat{\rho}(\boldsymbol{q}_1+\cdots +\boldsymbol{q}_d)
    + O(q^{d+1}),
    \label{4d density b}
\end{align}
where 
$\rho_0=
N_-/\Omega =
\mbox{(the number of occupied bands)}/
\mbox{(the volume of the unit cell)}$ 
is the average density,
${\it Ch}_n$ is the $n$-th Chern number ($n=d/2$), 
\begin{align}
  {\it Ch}_{n}
  &:= \frac{1}{n!}\left( \frac{\ii}{2\pi} \right)^n
    \int_{\mathrm{BZ}} \mathrm{Tr}_{{\it occ}} (\mathcal{F}^n)
    \nonumber \\
  &=
\frac{1}{(d/2)!} 
    \left( \frac{\ii}{2\pi} \right)^{d/2}
    \left(
    \frac{1}{2}
    \right)^{d/2}
    \int_{\mathrm{BZ}}
    \dd^d \boldsymbol{k}\,
    \epsilon^{\mu_1\cdots \mu_d}\,
    \mathrm{Tr}_{\mathrm{\it occ}}
    (\mathcal{F}_{\mu_1\mu_2}
    \cdots
    \mathcal{F}_{\mu_{d-1}\mu_{d}}),
\end{align}
where the trace is taken over the 
occupied band indices. 
We note that,
with this definition of 
the Chern numbers, 
the level of the Chern-Simons
term in the effective response action 
of a $(d+1)$-dimensional topological insulator
coincides with
${\it Ch}_{d/2}$
(see, e.g., 
\cite{Fukaya_2020}).
In terms of 
the envelope functions $f_{\alpha}(\boldsymbol{x})$
the bulk algebra \eqref{4d density b}
translates into
\begin{align}
[\hat{\rho}(f_1),  \cdots, \hat{\rho}(f_d)
]_{{\it mod}}
\sim 
 \rho^{-1}_0 
(d/2)!
    \left(\frac{\ii}{2\pi} \right)^{{d}/{2}}
    {\it Ch}_{{d}/{2}}\,
    \hat{\rho}(
    \{f_1,\cdots,f_d\}).
    \label{main result 2}
\end{align}
Setting
$f_{\alpha}(\boldsymbol{x})
\to
f_{\alpha}({\bf x})g(x_d)$
and replacing $\hat{\rho}\to \rho_0$
deep inside the bulk, 
we deduce the boundary algebra, 
\begin{align}
  &
    \big[ \hat{\varrho}(f_1),
    \cdots,
    \hat{\varrho}(f_d)
    \big]_{{mod}}
    \sim
    -
    (d/2)!
    \left(
    \frac{\ii}{2\pi}
    \right)^{d/2}
    {\it Ch}_{d/2}
    \,
    \int_{\partial M_{d}}
    {f}_1
    \dd {f}_2 
    \cdots
    \dd {f}_d, 
\end{align}
where we noted
$\int^{\infty}_{-\infty} \dd x_d\,
{\partial (g^{d})}/{\partial x_d}
\rho(\boldsymbol{r})
\sim
\rho_0
    \int^{0}_{-\infty} \dd x_d\,
{\partial (g^{d})}/{\partial x_d}
$.
The bulk algebra \eqref{4d density b}
thus compares well with the boundary algebra \eqref{S40 3}.

The bulk algebra \eqref{4d density b}
is calculated using the Bloch wave functions. 
Instead,
we can work with the hybrid Wannier functions.
This in particular 
allows us to obtain 
the boundary (as well as bulk) density operator algebras. 
The use of the Wannier functions 
is discussed in Section 
\ref{The Wannier functions and the bulk-boundary relation}, 
by taking the (2+1)d and (4+1)d LLLs as an example. 

Finally, we conclude in Section \ref{discussion} by presenting the summary and outlook.  

\section{Second quantization, higher-brackets and noncommutative geometry}
\label{n-bracket and second quantization}

\subsection{Second quantization map and normal order}
\label{sec:2ndquantizationmap}

We start by introducing some notations.
Let $A$ be an operator on the single-particle Hilbert space $V$.
For basis $\{\ket{n}\}$ of $V$,
matrix elements of $A$ are given by $A_{mn} = \braket{m|A|n}$.
In the second quantization,
we consider the operator corresponding to $A$
(the second quantization map)
on the Fock space,
\begin{align}
\hat{Q}_0(A) = \sum_{mn} A_{mn} \hat{\psi}_m^{\dag} \hat{\psi}^{\ }_n,
\end{align}
where
$\hat{\psi}^{\dag}_n/\hat{\psi}^{\ }_n$ are the fermion creation/annihilation operator
associated to the single-particle state $|n\rangle$ satisfying canonical anticommutator relations, $\{\hat{\psi}_n,\hat{\psi}_m^\dag\}=\delta_{nm}$ and $\{\hat{\psi}_n, \hat{\psi}_m\}=0$.
For the product of two second-quantized operators
$\hat{Q}_0(A)$ and $\hat{Q}_0(B)$,
\begin{align}
	\hat{Q}_0(A)\hat{Q}_0(B)
	=
	\hat{Q}_0(AB)
	-
	\hat{Q}_{0,2} (A\otimes B)
	\label{Q0AQ0B}
\end{align}
where
\begin{align}
\hat{Q}_{0,2}(A\otimes B) = \sum_{klmn} A_{kl}B_{mn} \hat{\psi}^{\dag}_k \hat{\psi}^{\dag}_m
\hat{\psi}^{\ }_l 
\hat{\psi}^{\ }_n
\label{Q2}
\end{align}
is the second quantized map for a 2-particle operator $A \otimes B$.
Noting $\hat{Q}_{0,2}(A\otimes B)=\hat{Q}_{0,2}(B\otimes A)$,
the commutator
$[\hat{Q}_0(A), \hat{Q}_0(B)]$ is given by
\begin{align}
	[\hat{Q}_0(A),\hat{Q}_0(B)] =
	\hat{Q}_0([A,B])
	\label{Q0Q0} 
	.
\end{align}

Let us consider a many-body ground state in the Fock space,
which is a Fermi-Dirac sea given by
$\ket{GS} = \prod_{n<0} \hat{\psi}^{\dag}_n \ket{0}$.
For the specified ground state,
$\hat{Q}_0(A)$ may be ill-defined
since
the expectation value 
$\langle {\it GS}| \hat{Q}_0(A) |{\it GS}\rangle$
may be divergent.
This can be resolved for many operators $A$ by
introducing a normal-ordered counterpart of $\hat{Q}_0(A)$:
\begin{align}
\hat{Q}(A) = \sum_{mn} A_{mn} \normord{ 
\hat{\psi}_m^{\dag} \hat{\psi}^{\ }_n } ,
\end{align}
where we define the normal order as
\begin{align}
\normord{\hat{\psi}^{\dag}_m \hat{\psi}^{\ }_n} = \left\{ \begin{array}{ll}
- \hat{\psi}^{\ }_n 
\hat{\psi}^{\dag}_m & (n=m<0), \\
\hat{\psi}^{\dag}_m \hat{\psi}^{\ }_n & (\mathrm{otherwise}). \\
\end{array} \right.
\end{align}
For later use, let us introduce a notation
\begin{align}
A_{\epsilon \epsilon'} = P_{\epsilon} A P_{\epsilon'} \quad (\epsilon,\epsilon' = \pm),
\end{align}
where $P_{-}$ ($P_{+}$) is the projection onto $n<0$ ($n>0$) states. We also define the operator $F=P_+-P_-$. This is a self-adjoint operator squaring to the identity operator.  

For the product of two normal-ordered operators
$\hat{Q}(A)$ and $\hat{Q}(B)$,
\begin{align}
&
\hat{Q}(A)\hat{Q}(B)
\nonumber \\
&\quad =
\hat{Q}_{0,2}(A\otimes B)
-
\hat{Q}_0(A) \mathrm{Tr}\, (B_{--})
-
\mathrm{Tr}\, (A_{--}) 
\hat{Q}_0(B)
+
\mathrm{Tr} \big( (AB)_{--} \big)
\nonumber \\
&\quad =
\hat{Q}_{2}(A\otimes B)
+
\hat{Q}( AP_+B - BP_-A)
+
\mathrm{Tr}\, (A_{-+}B_{+-}),
\end{align}
where $\hat{Q}_{2}$ is the normal-ordered counterpart of $\hat{Q}_{0,2}$,
\begin{align}
\hat{Q}_{2}(C) := \sum_{klmn} C_{klmn}
\normord{
\hat{\psi}^{\dag}_k 
\hat{\psi}^{\ }_l 
\hat{\psi}^{\dag}_m
\hat{\psi}^{\ }_n}.
\end{align}
Here,
normal ordering for 
$n$-body operators 
with $n>1$
is defined 
analogously to the 1-body case;
all fermion operators 
annihilating
the ground state,
i.e., $\hat{\psi}_{m>0}$ and $\hat{\psi}^{\dag}_{m<0}$,
are moved to the right. 
E.g., 
$
\normord{
\hat{\psi}^\dag_k 
\hat{\psi}^{\ }_l 
\hat{\psi}^\dag_m 
\hat{\psi}^{\ }_n}
$
is
$
\hat{\psi}^\dag_k \, 
\normord{ \hat{\psi}^\dag_m  \hat{\psi}^{\ }_n } \,
\hat{\psi}^{\ }_l$
if $k,l>0$,
$
-\hat{\psi}^{\ }_l
\normord{ \hat{\psi}^\dag_m \hat{\psi}^{\ }_n }\, 
\hat{\psi}^\dag_k$
if $k,l<0$, 
$
\hat{\psi}^\dag_k \hat{\psi}^{\ }_l 
\normord{ \hat{\psi}^\dag_m  \hat{\psi}^{\ }_n }$
if $k>0, l<0$, 
and
$
\normord{ \hat{\psi}^\dag_m  \hat{\psi}^{\ }_n } \,
\hat{\psi}^\dag_k \hat{\psi}^{\ }_l
$
if $k<0, l>0$. 
Noting $\hat{Q}_{2}(A\otimes B)=
\hat{Q}_{2}(B\otimes A)$,
the commutator of
two normal-ordered operators
$\hat{Q}(A)$ and $\hat{Q}(B)$
is given by
\begin{align}
&
[\hat{Q}(A), \hat{Q}(B)] = \hat{Q}([A,B]) + b_2([A,B]),
\label{2-bracket}
\end{align}
where $b_2([A,B])$ is a c-number and defined by 
\begin{align}
b_2([A,B]) = \mathrm{Tr}\,([A,B]_{--}).
\end{align}

For many operators of interest in physics, $b_2([A, B])$ is not well-defined. 
However, one can make it well-defined by the following computation, 
\begin{align}
b_2([A,B]) &= \mathrm{Tr}\big((AB)_{--} - (BA)_{--} \big)
\nonumber \\
&= \mathrm{Tr}\, (A_{-+}B_{+-}-B_{-+}A_{+-}) +  \mathrm{Tr}\, ([A_{--},B_{--}]) 
\end{align}
and dropping the term $\mathrm{Tr}\, ([A_{--},B_{--}])$.
We emphasize that, since this is done for operators where neither
$\mathrm{Tr} \big((AB)_{--}\big)$ nor
$\mathrm{Tr} \big((BA)_{--}\big)$
are well-defined, setting $\mathrm{Tr}\, ([A_{--}, B_{--}])$ to zero can only be done after a nontrivial regularization in general (since for operators in infinite dimensions, the trace of a commutator is not well-defined in general).
Thus, by some regularization, one can replace $b_2([A,B])$ in Eq.\ \eqref{2-bracket} by 
\begin{equation} 
S_2(A,B)=\mathrm{Tr}\, (A_{-+}B_{+-}-B_{-+}A_{+-}). 
\label{S2} 
\end{equation} 
The resulting abstract current algebra,  
\begin{align}
&
[\hat{Q}(A), \hat{Q}(B)] = \hat{Q}([A,B]) + S_2(A,B), 
\label{2-bracket alg}
\end{align}
has many special cases of interest in physics. 
Specifically, the formula \eqref{2-bracket alg}
can be used for operators appearing in (1+1)d quantum field theories,
leading to, for example, the (1+1)d current algebra, 
the affine Kac-Moody algebra, Virasoro algebra, $W_{1+\infty}$ algebra
\cite{Lundberg:1976nu,Pressley1986loop, Mickelsson:1987ns}.

We conclude this section with remarks. 
First, we note that in intermediate steps of the derivation of Eq.\ \eqref{2-bracket alg} above, 
it is understood that the operators $A,B$ are replaced by regularized operators $A^\Lambda,B^\Lambda$, with $\Lambda$ some regularization parameter, so that all traces are well-defined. It is clear that the final result (after removing the regularization) does not depend on what regularization is used, which is why we do not specify the regularization. 
To give a specific example, 
consider the case when $V$ is the single-particle Hilbert space of square-integrable functions on $M_{1}$,
and $A$ is a real-valued function 
$A=f(x)$.
In Fourier space, $f(x)$ is represented 
by its Fourier counterpart, 
$\tilde{f}(p-p') = \int_{M_1}dx\, f(x) 
e^{ \ii (p-p')\cdot x}$.
A natural operator regulation of this would be to replace $\tilde{f}(p-p')$ 
by 
$
\tilde{f}^{\Lambda}(p,p')
=
\chi(p/\Lambda) \tilde{f}(p-p') \chi(p'/\Lambda)
$,
where $\chi(\xi)$ is
some function of $\xi\ge 0$ which is smooth at $\xi=0$
such that 
$\chi(0)=1$ and $\chi(\xi)\to 0$
as $\xi \to \infty$,
for example, 
$\chi(\xi)= e^{-\xi^2}$
or 
$\chi(\xi) = 1/(\xi^2+1)$.
The trace of 
commutators of regularized operators 
is always 0, and hence 
$b_2([A,B])
:=
\lim_{\Lambda\to \infty}\, 
\mathrm{Tr}\, ( [A^{\Lambda},B^{\Lambda}]_{--})
=
S_2(A,B)$.
Here, $S_2(A,B)$ 
is given in terms of traces 
of trace-class operators
and hence
independent of the regularization.
\footnote{
In addition to the operator regularization mentioned in the main text, in which one replaces 
$A$ by $A^{\Lambda}$,
there is another so-called vacuum regularization.
In this regularization,
we introduce a regularized vacuum 
$|{\it GS}_{\Lambda} \rangle$
as the state where only the eigenstates with energies in the range $-\Lambda<E<0$ are filled. Clearly, 
$|{\it GS}_{\Lambda}\rangle 
\to |{\it GS}\rangle$.
With the regularized vacuum,  
$b_2([A,B]) = \mathrm{Tr}\, (P_- [A,B])$
can be regularized as 
$b^{\Lambda}_2([A,B]) = \mathrm{Tr}\, (P_{\Lambda} [A,B])$
with the cut-off trace
$
 \mathrm{Tr}_\Lambda(A):=  \mathrm{Tr}(P_\Lambda A) 
$
where $P_\Lambda$ is the projection to subspace of energies in the range $-\Lambda<E<0$. 
When $[A,B]=0$, e.g., 
when we consider scalar functions $A=f_1$ and $B=f_2$,  
it therefore is clear that $b^\Lambda_2([f_1,f_2])= \mathrm{Tr}_\Lambda(P_- [f_1,f_2])=0$.
Therefore,
if one uses the vacuum regularization, 
one has to renormalize the commutator of the fermion currents by subtracting $R_2= \mathrm{Tr}\, ([A_{--}, B_{--}])$ to obtain the correct result. 
The same reasoning applies to all $d=2,3,6,\cdots$; 
In the Abelian case, $[f_1,\ldots,f_d]=0$, and it therefore is clear that $b^\Lambda_d([f_1,\ldots,f_d])= \mathrm{Tr}_\Lambda(P_- [f_1,\ldots,f_d])=0$; one can show that the same is true also in the non-Abelian case. 
Thus, for the vacuum regularization, $R_d^{\text{vac-reg}}=-S_d$ for all $d=2,4,6,\ldots$. 
Following the 2d case,
our proposal is to renormalize the $d$-bracket of densities in higher dimensions by removing $R_d$.
%
}

Second, we stress that it is important that the above computation is done for non-commuting operators $A,B$ even if one is only interested in cases where $[A,B]=0$: the regularized operators $A^\Lambda,B^\Lambda$ do not commute even if $A,B$ do. Moreover, allowing for non-commuting operators makes clear that there is an important consistency condition: on the one-particle level, the commutator satisfies the Jacobi identity
\begin{align} 
\frac12\epsilon^{\alpha\beta\gamma}[[A_\alpha,A_\beta],A_\gamma]=0, 
\end{align} 
and Eq.\ \eqref{2-bracket alg} is only consistent if the operators $\hat{Q}(A)$ satisfy the Jacobi identity as well. 
Clearly, this is true if and only if  
\begin{align} 
\delta S_2(A_1,A_2,A_3):= \frac12\epsilon^{\alpha\beta\gamma}S_2([A_\alpha,A_\beta],A_\gamma)=0, 
\label{deltaS2} 
\end{align} 
which is known as cocycle condition. It is well-known that this condition is satisfied for the Schwinger term in Eq.\ \eqref{S2}.

\subsection{Relation to non-commutative geometry}\label{sec:NCG} 
We shortly describe a mathematical interpretation of the Schwinger term Eq.\ \eqref{S2} in the context of noncommutative geometry
\cite{Connes1994,Gracia2001} which we use as a guide in Section~\ref{sec:nbrackets} to generalize 
Eq.\ \eqref{2-bracket alg} to 3+1 dimensions. 
For that, it is useful to write operators as $2\times 2$-matrices as follows
\cite{Carey1987},
$$
A=\left(\begin{array}{cc} A_{++} & A_{+-} \\ A_{-+} & A_{--} \end{array}\right) 
$$
so that $F$ can be identified with the Pauli $\sigma_3$-matrix. Thus, 
$$
[F,A]= 2\left(\begin{array}{cc} 0  & A_{+-} \\ -A_{-+} & 0 \end{array}\right) , \quad \{F,A\}= 2\left(\begin{array}{cc} A_{++} & 0 \\ 0 & -A_{--} \end{array}\right), 
$$
allowing us to write the Schwinger term in Eq.\ \eqref{S2} as  
\begin{equation} 
S_2(A,B)=\frac12 \mathrm{Tr}_c\, (A[F,B]), 
\label{S2v2} 
\end{equation} 
with the conditional trace defined as follows,
\begin{equation*} 
\mathrm{Tr}_c\, (A):= \mathrm{Tr}\, (A_{++} + A_{--}). 
\end{equation*} 
Note that $\mathrm{Tr}_c$ is a generalization of the conventional trace
(since, for the conventional trace $\mathrm{Tr}\,(A)$ to exist, all operators
$A_{\epsilon\epsilon'}$
for $\epsilon,\epsilon'=\pm$ need to be trace class,
whereas only $A_{++}+A_{--}$ needs to be trace class for $\mathrm{Tr}_c\, (A)$ to exist).  It is also interesting to note that the Schwinger terms $S_2$ can be written using the conventional trace and the 2d epsilon symbol as follows, 
\begin{equation} 
S_2(A_1,A_2)=\frac18 \epsilon^{\alpha\beta}\mathrm{Tr}\, (F[F,A_\alpha][F,A_\beta]).  
\end{equation}

In noncommutative geometry, $\ii A[F,B]$ is a natural generalization of the de Rham 1-form $fdg$, and $\mathrm{Tr}_c$ is a corresponding generalization of integration of de Rham forms
\cite{Langmann1995}.
This is made precise in 2d by the identity 
\begin{align} 
\mathrm{Tr}_c(f_1[F,f_2]) =-\frac{\ii}\pi \int_{M_1} f_1\dd{f}_2 
\label{Trc1+1d}
\end{align} 
with $f_1\dd{f}_2= f_1\frac{\partial f_2}{\partial x}\dd{x}$,
where $f_{\alpha}=f_\alpha(x)$ are real-valued functions on a 1-dimensional space $M_1$ (which can be either $\mathbb{R}$ (real line) or $\mathbb{S}^1$ (circle)), and $F$ is the sign of the 1d Weyl operator $-\ii \partial_x$ (i.e., $F$ amounts to multiplication with $\mathrm{sign}(p)$ in Fourier space). By specializing Eqs.\ \eqref{2-bracket alg} and \eqref{S2v2} to the one-particle Hilbert space of square-integrable functions on $M_1$ (so that $f_\alpha$ and $F$ can be identified with operators on this Hilbert space) and using the identity in Eq.\ \eqref{Trc1+1d}, one obtains 
$$
[\hat{Q}(f_1),\hat{Q}(f_2)] = -\frac{\ii}{2\pi} \int_{M_1} f_1\dd{f}_2 , 
$$ 
and by  identifying $\hat{Q}(f_i)$ with $\hat{\varrho}(f_i)$ this becomes exactly the (1+1)d current algebra in Eq.\ \eqref{bdry algebra} for $\nu=1$ and $\partial M_2=M_1$.

It is known that there is a natural generalization of Eq.\ \eqref{Trc1+1d} to arbitrary 
odd space dimensions $d-1=2n-1$
\cite{Langmann1995}, 
\begin{align} 
\mathrm{Tr}_c\, (f_1[F,f_2]\cdots [F,f_{d}]) =  \left(-\frac{\ii}{2\pi}\right)^{d/2} \frac{2^{d} (d/2)!}{d!}\int_{M_{d-1}} f_1\dd{f}_2\cdots \dd{f}_{d} 
\label{Trc2nd}
\end{align} 
(note that the constant in front of the integral is written as $(-\ii)^{d-1}(2\ii)^{d/2-1}
2\pi^{(d-1)/2}/(d-1)(2\pi)^{d-1}\Gamma( (d-1)/2)$ in Ref.~\cite{Langmann1995}) with the following de Rham form in $(d-1)$ space dimensions, 
\begin{align*}
f_1\dd{f}_2\cdots \dd{f}_{d} := 
\epsilon^{i_1\cdots i_{d-1}} f_1\frac{\partial f_2}{\partial x_{i_1}}\cdots \frac{\partial f_{d}}{\partial x_{i_{d-1}}}\dd{x}_1\cdots \dd{x}_{d-1} 
\end{align*} 
for  $f_{\alpha}=f_{\alpha}(\bf{x})$ real-valued functions on ($d-1$)-dimensional
space $M_{d-1}$, which can be either $\mathbb{R}^{d-1}$ 
or the torus $\mathbb{T}^{d-1}$.
Here, as the grading operator $F$,
we consider the sign of the $(d-1)$d Weyl operator
$-\ii\sum_{i=1}^{d-1}\Gamma_{i}\frac{\partial}{\partial x_i}$ with $2^{d/2-1}\times
2^{d/2-1}$-matrices $\Gamma_i$ such that $\{\Gamma_{i},\Gamma_{j}\} = 2\delta_{ij}$
(i.e., $F$ amounts to multiplication with
$\sum_{i=1}^{d-1} \Gamma_{i} p_i /|\mathbf{p}|$ in Fourier space);
in particular, in (3+1)d, 
\begin{align} 
\mathrm{Tr}_c\, (f_1[F,f_2][F,f_3])[F,f_4]) =  -\frac{1}{3\pi^2}\int_{M_3} f_1\dd{f}_2\dd{f}_3\dd{f}_4 
\label{Trc3+1d}
\end{align} 
where $\Gamma_i=\sigma_i$ are the Pauli sigma matrices.

\subsection{Higher brackets}
\label{sec:nbrackets}
In Section~\ref{sec:2ndquantizationmap}, we have encountered the Schwinger term $S_2(A,B)$ providing a $U(1)$ central extension term of a Lie algebra; 
see Ref.\ \cite{Langmann:1996sg} for further details. 
To generalize this result for topological insulators in higher dimensions,
we will now consider the products of more than two second-quantized operators.
In general, for second quantized operators $\hat{O}_{\alpha=1,\ldots,d}$, 
we consider the $d$-bracket defined by 
\begin{align}
[\hat{O}_1, \hat{O}_2 ,\dots, \hat{O}_{d}] =
\epsilon^{\alpha_1 \alpha_2\cdots \alpha_d} 
\hat{O}_{\alpha_1} \hat{O}_{\alpha_2} \cdots 
\hat{O}_{\alpha_{d}}.
\end{align}
Of special interest to us is the case where $\hat{O}_\alpha$ is given
by the second quantization map 
of some single-particle operator $A_\alpha$; 
namely, we are interested in the $d$-bracket of the unordered second quantization maps
$\hat{Q}_0(A_\alpha)$
and their normal-ordered counterparts $\hat{Q}(A_\alpha)$, 
i.e.,
$[\hat{Q}_0(A_1),\hat{Q}_0(A_2), \dots, \hat{Q}_0(A_{d})]$
and
$[\hat{Q}(A_1),\hat{Q}(A_2), \dots, \hat{Q}(A_{d})]$ for even $d$.

In the rest of this section, we focus on $d=4$ (generalizations to arbitrary even-$d$ brackets can be found in Section~\ref{sec:higherdim}).
First, for unordered operators
$\hat{Q}_0(A_{\alpha})$,
\begin{align}
&
[\hat{Q}_0(A_1), \hat{Q}_0(A_2),
\hat{Q}_0(A_3), \hat{Q}_0(A_4)]
\nonumber \\
&= \epsilon^{\alpha\beta\gamma\delta}
\hat{Q}_0(A_{\alpha}) \hat{Q}_0(A_{\beta}) 
\hat{Q}_0(A_{\gamma}) \hat{Q}_0(A_{\delta})
\nonumber \\
&= \hat{Q}_0([A_1,A_2,A_3,A_4])
+ \epsilon^{\alpha\beta\gamma\delta} \hat{Q}_{0,2}
(A_{\alpha}A_{\beta}\otimes A_{\gamma}A_{\delta}),
\label{regular 4-bracket a}
\end{align}
where we noted Eq.\ \eqref{Q0AQ0B} and
\begin{align}
&
[\hat{Q}_0(A_1), 
\hat{Q}_0(A_2),
\hat{Q}_0(A_3), 
\hat{Q}_0(A_4)]
\nonumber \\
&=
\frac{1}{8} \epsilon^{\alpha\beta\gamma\delta} \Big\{
\hat{Q}_0([A_\alpha, A_\beta]),\,
\hat{Q}_0([A_\gamma, A_\delta])
\Big\}
\nonumber \\
&=
\frac{1}{8} \epsilon^{\alpha\beta\gamma\delta}
\Big[
\hat{Q}_0(\{ [A_\alpha, A_\beta], 
[A_\gamma, A_\delta]\})
+
2 \hat{Q}_{0,2} 
([A_\alpha, A_\beta]
\otimes [A_\gamma,A_\delta])
\Big].
\end{align}
This suggests to define a {\it modified} 4-bracket by subtracting $\epsilon^{\alpha\beta\gamma\delta} \hat{Q}_{0,2}(A_{\alpha}A_{\beta}\otimes A_{\gamma}A_{\delta})$
from the four-brackets $[\hat{Q}_0(A_1), \hat{Q}_0(A_2),\hat{Q}_0(A_3), \hat{Q}_0(A_4)]$, 
\begin{align} 
[\hat{Q}_0(A_1), &\hat{Q}_0(A_2),\hat{Q}_0(A_3),\hat{Q}_0(A_4)]_{mod} \nonumber \\ & := [\hat{Q}_0(A_1), \hat{Q}_0(A_2),\hat{Q}_0(A_3), \hat{Q}_0(A_4)] - 
\epsilon^{\alpha\beta\gamma\delta} \hat{Q}_{0,2}(A_{\alpha}A_{\beta}\otimes A_{\gamma}A_{\delta}).
\end{align} 
Indeed, this yields 
\begin{equation} 
[\hat{Q}_0(A_1), \hat{Q}_0(A_2),\hat{Q}_0(A_3), \hat{Q}_0(A_4)]_{mod} = \hat{Q}_0([A_1,A_2,A_3,A_4]), 
\end{equation} 
which is a natural analogue of 
Eq.\ \eqref{Q0Q0}. 

Similarly, the  $4$-bracket of the normal ordered operators $\hat{Q}(A_\alpha)$ is evaluated as
\begin{align}
&
[\hat{Q}(A_1), 
\hat{Q}(A_2),
\hat{Q}(A_3), 
\hat{Q}(A_4)]
\nonumber \\
&=
\frac{1}{8} \epsilon^{\alpha\beta\gamma\delta} \Big\{
\hat{Q}_0([A_\alpha, A_\beta]),\,
\hat{Q}_0([A_\gamma, A_\delta])
\Big\}
\nonumber \\
& = \hat{Q}([A_1,A_2,A_3,A_4]) + b_4([A_1,A_2,A_3,A_4]) + \epsilon^{\alpha\beta\gamma\delta}\hat{Q}_{0,2} 
(A_\alpha A_\beta \otimes A_\gamma A_\delta)
\end{align}
where the c-number 
\begin{align}
b_4([A_1,A_2,A_3,A_4])
&= \mathrm{Tr}\, \big( [A_1,A_2,A_3,A_4]_{--} \big)
\label{def b4}
\end{align}
arises from normal ordering, as before. 
This c-number piece can be conveniently split into two parts,
$b_4=R_4+S_4$, where (to simplify notation, we sometimes write $A,B,C,D$ instead
of $A_1,A_2,A_3,A_4$ in the following) 
\begin{align}
R_4(A,B,C,D)
&= \epsilon^{ABCD} \mathrm{Tr}\, \big(A_{-+} B_{++} C_{++}D_{+-}+A_{--} B_{--} C_{-+}D_{+-}\big), 
\nonumber \\
S_4(A,B,C,D)
&= \epsilon^{ABCD} \mathrm{Tr}\, \big(A_{-+}B_{+-}C_{-+}D_{+-}\big).
\label{R4S4}
\end{align}
This can be seen by computing 
\begin{align}
b_4([A,B,C,D])
&= \epsilon^{ABCD} \sum_{\epsilon, \epsilon, \epsilon''
= \pm}\mathrm{Tr} \big(A_{- \epsilon} B_{\epsilon \epsilon'} C_{\epsilon' \epsilon''} D_{\epsilon'' -}\big)
\nonumber \\
&= \epsilon^{ABCD} \mathrm{Tr} \big(A_{-+}B_{+-}C_{-+}D_{+-}+A_{-+} B_{++} C_{++}D_{+-}+A_{--} B_{--} C_{-+}D_{+-}\big), 
\label{b4computation} 
\end{align}
using the cyclicity of trace and the antisymmetry for labels $\{A, B, C, D\}$ to verify that  
\begin{align}
\epsilon^{ABCD} \mathrm{Tr}\big(A_{--}B_{--}C_{--}D_{--}\big)
= \frac12 \epsilon^{ABCD} \mathrm{Tr} \big([A_{--},B_{--}C_{--}D_{--}]\big) = 0  
\end{align}
and 
\begin{align} 
  &
\epsilon^{ABCD} \mathrm{Tr} \big(A_{--}B_{-+}C_{+-}D_{--} + A_{-+}B_{+-}C_{--}D_{--} + A_{--}B_{-+}C_{++}D_{+-} + A_{-+}B_{++}C_{+-}D_{--}  \big) 
  \nonumber \\
  &\quad
= \epsilon^{ABCD} \mathrm{Tr} \big( [A_{--},B_{-+}C_{+-}D_{--}] + [A_{--},B_{-+}C_{++}D_{+-} ]\big) = 0.   
\end{align} 
Using the grading operator $F$,
it is straightforward to write Eq.\ \eqref{R4S4} using notation introduced in Section~\ref{sec:NCG}, 
\begin{align}
R_4(A_1,A_2,A_3,A_4)
  & = \frac{1}{16}
  \epsilon^{\alpha\beta\gamma\delta}
    \mathrm{Tr}\, (F\{F, A_\alpha\}\{F, A_\beta\} [F,A_\gamma] [F,A_\delta]),
  \nonumber  \\
S_4(A_1,A_2,A_3,A_4)
  &= -\frac{1}{32}
    \epsilon^{\alpha\beta\gamma\delta} \mathrm{Tr}\,(
      F[F,A_\alpha] [F,A_\beta] [F,A_\gamma] [F,A_\delta]).
    \label{R4S41}
\end{align}
Moreover, the latter formula is equivalent to
\cite{Langmann1995} 
\begin{equation} 
  S_4(A_1,A_2,A_3,A_4) = -\frac{1}{16}\epsilon^{\alpha\beta\gamma\delta} \mathrm{Tr}_c\,
  \big(A_\alpha [F,A_\beta] [F,A_\gamma] [F,A_\delta]\big), 
\label{S4}
\end{equation} 
which makes manifest that $S_4$ is a generalization to noncommutative geometry (NCG) of an integral of a de Rham form in (3+1)d. 

Our discussion so far is rather general and applies to any system 
with the fermion Fock space endowed with a grading operator. 
We now specialize 
to the Hilbert space of square-integrable functions on 3d space $M_3$ 
with spin 1/2 degrees of freedom
and $F$ equal to the sign of the 3d Weyl operator 
$-\ii\sum_{i=1}^3 \sigma_i\frac{\partial}{\partial x_i}$.
As for $A_{\alpha}$, we consider 
spin-independent,
real-valued functions on $M_{d-1}$,
$f_{\alpha}({\bf x})$.
Here, remarks like the ones in the final paragraph of
Section~\ref{sec:NCG} apply. 
In particular, the precise meaning of 
the above expressions
is to replace the
operators $A_\alpha$ by 
their regularized counterparts $A_\alpha^\Lambda$ and
then take a limit where the regularization parameter $\Lambda$ is removed.
(We note that the cyclicity of 
trace, used in deriving above expressions,
applies to the product of 
regularized operators.)
Now, the rationale for the splitting 
of $b_4$ into $S_4$ and $R_4$ is that 
$S_4$ is given
in terms of traces of trace-class operators,
while $R_4$ is not:
As pointed out by Mickelsson and Rajeev 
\cite{Mickelsson:1987ns}, in 
$d-1$ spatial dimensions, 
the pertinent operators 
$A_{\alpha}$ and $F$ obey the condition
that
$[F, A_{\alpha}]$,
and hence $A_{\alpha,+-}$ and $A_{\alpha,-+}$,
are 
the $d$-th Schatten class operators on 
the single-particle Hilbert space.
Here, for an operator $A$ in the $d$-th Schatten class,  
$(A^{\dag}A)^{d/2}$ are trace-class.
In particular,
for $d=4$
products like 
$
A_{\alpha +-}
A_{\beta -+} 
A_{\gamma +-}A_{\delta -+}
$
are trace-class (see 
\cite{alma991008783459703276}
for mathematical background on Schatten classes).
Consequently, 
while
$S_4$ does not depend on the regularization used, $R_4$ is affected by the regularization.
For example,
if one uses a naive cutoff scheme,
in which one cut-offs the matrix elements of 
$f_{\alpha}$ 
in momentum space,
$\tilde f^\Lambda(\mathbf{p},\mathbf{q})=\chi(|\mathbf{p}|/\Lambda)\tilde f(\mathbf{p}-\mathbf{q})\chi(|\mathbf{q}|/\Lambda)$ for some cutoff function like $\chi(x)=e^{-x^2}$,  
one can show by a brute-force calculation,
$R_4=-S_4$ and hence $b_4=0$.
While 
the vanishing of $b_4$ may naively be consistent with 
an expression like Eq.\ \eqref{def b4},
it is not 
robust since in other regularization
schemes, e.g., spin-dependent ones,
one can get $R_4 \neq -S_4$ \cite{LangmannUnpublished}.
%

Since $S_4$, but not $R_4$, is well-defined for the pertinent one-particle operators in (3+1)d
\cite{Langmann1995}, we not only subtract $\epsilon^{\alpha\beta\gamma\delta}\hat{Q}_{0,2} 
(A_\alpha A_\beta \otimes A_\gamma A_\delta)$ but also $R_4(A_1,A_2,A_3,A_4)$ to obtain a well-defined modified 4-bracket of the normal-ordered operators, 
\begin{align} 
[\hat{Q}(A_1), &\hat{Q}(A_2),\hat{Q}(A_3),\hat{Q}(A_4)]_{mod} \nonumber \\ & := [\hat{Q}(A_1), \hat{Q}(A_2),\hat{Q}(A_3), \hat{Q}(A_4)] - 
\epsilon^{\alpha\beta\gamma\delta} \hat{Q}_{0,2}(A_{\alpha}A_{\beta}\otimes A_{\gamma}A_{\delta})-R_4(A_1,A_2,A_3,A_4); 
\label{modbracket}
\end{align} 
This yields
\begin{equation} 
[\hat{Q}(A_1), \hat{Q}(A_2),\hat{Q}(A_3),\hat{Q}(A_4)]_{mod} = \hat{Q}([A_1,A_2,A_3,A_4]) + S_4(A_1,A_2,A_3,A_4) . 
\label{4bracketrelation} 
\end{equation} 
%
For $f_\alpha$ real-valued functions on $M_3$, we can identify $\hat{Q}(f_\alpha)$ with the smeared density operators $\hat{\varrho}(f_\alpha)$,  and Eq.\ \eqref{Trc3+1d} implies 
\begin{align} 
[\hat{\varrho}(f_1),\hat{\varrho}(f_2),\hat{\varrho}(f_3),\hat{\varrho}(f_4)]_{mod} & = \frac{1}{2\pi^2}\int_{M_3} f_1 \dd{f}_2 \dd{f}_3 \dd{f}_4, 
\label{4currentalgebra} 
\end{align} 
using the identity
$
\int_{M_3} \epsilon^{\alpha\beta\gamma\delta} f_\alpha \dd{f}_\beta \dd{f}_\gamma \dd{f}_\delta =4! \int_{M_3} f_1 \dd{f}_2 \dd{f}_3 \dd{f}_4. 
$
This is our chiral current algebra in (3+1)d and one of the main results in this paper.

As we will see later in Section 
\ref{Bulk density operator algebra},
the robust part of $b_4$, 
i.e., $S_4$, compares well with the bulk calculation, consistent with 
the bulk-boundary corresponcence.
Our proposal 
to remove (renormalize) $R_4$
is further motivated by 
mathematical consistency: 
It is known that the 4-bracket of
one-particle operators satisfies the following generalized Jacobi identity
\cite{de_Azc_rraga_1997},
\begin{align} 
\frac1{144}\epsilon^{\alpha_1\alpha_2\alpha_3\alpha_4\alpha_5\alpha_6\alpha_7}[[A_{\alpha_1},A_{\alpha_2},A_{\alpha_3},A_{\alpha_4}],A_{\alpha_5},A_{\alpha_6},A_{\alpha_7}]=0 , 
\label{GJI} 
\end{align} 
and it is therefore natural to request that Eq.\ \eqref{4bracketrelation}
is consistent in the sense that 
\begin{align*} 
\frac1{144}\epsilon^{\alpha_1\alpha_2\alpha_3\alpha_4\alpha_5\alpha_6\alpha_7}[[\hat{Q}(A_{\alpha_1}),\hat{Q}(A_{\alpha_2}),\hat{Q}(A_{\alpha_3}),\hat{Q}(A_{\alpha_4})]_{mod},\hat{Q}(A_{\alpha_5}),\hat{Q}(A_{\alpha_6}),\hat{Q}(A_{\alpha_7})]_{mod}=0, 
\end{align*} 
 similarly as for the commutator relations discussed in Section~\ref{sec:2ndquantizationmap}. 
Clearly, this consistency is fulfilled for operators $A_\alpha$ satisfying the cocycle condition $\delta S_4(A_1,\ldots,A_7) =0$ where 
\begin{equation} 
\delta S_4(A_1,\ldots,A_7) := \frac1{144}\epsilon^{\alpha_1\alpha_2\alpha_3\alpha_4\alpha_5\alpha_6\alpha_7}S_4([A_{\alpha_1},A_{\alpha_2},A_{\alpha_3},A_{\alpha_4}],A_{\alpha_5},A_{\alpha_6},A_{\alpha_7}) . 
\end{equation} 
If we specialize to real-valued function, $A_\alpha=f_\alpha$, it is clear from  $S_4(f_1,\ldots,f_4)\propto \int_{M_3} f_1 \dd{f}_2 \dd{f}_3 \dd{f}_4$  and $[f_1,f_2,f_3,f_4]=0$ that 
\begin{equation} 
\delta S_4(f_1,\ldots,f_7)=0, 
\label{deltaS4fi} 
\end{equation} 
i.e., the (3+1)d current algebra above is consistent. 

We stress that the consistency condition $\delta S_4=0$ does not have to be fulfilled for regularized operators: 
Eq.\ \eqref{deltaS4fi}
is enough to prove the consistency of our (3+1)d current algebra. Still,  it is interesting to know if Eq.\ \eqref{4bracketrelation} is consistent for other special cases of interest in physics.  
We therefore computed $\delta S_4(A_1,\ldots,A_7)$ directly and obtained the following result
\cite{LangmannUnpublished}, 
\begin{align} 
  \delta S_4(A_1,\ldots,A_7) =-\frac{1}{768}
  \epsilon^{\alpha_1\alpha_2\alpha_3\alpha_4\alpha_5\alpha_6\alpha_7}
  \mathrm{Tr}\, \big(A_{\alpha_1}\{F,A_{\alpha_2}\}[F,A_{\alpha_3}]
  [A_{\alpha_4},[F,A_{\alpha_5}][F,A_{\alpha_6}]]
  [F,A_{\alpha_7}]\big). 
\label{deltaS4} 
\end{align} 
Since this is not identically zero, Eq.\ \eqref{4bracketrelation} is not consistent for {\em all} operators such that 
Eq.\ \eqref{4bracketrelation} is well-defined. However, this result shows that Eq.\ \eqref{4bracketrelation} is consistent for another important case if interest in physics: as shown in Section~\ref{sec:nonAbelian}, 
Eq.\ \eqref{deltaS4} and known results in the literature
\cite{Langmann_1994} imply that Eq.\ \eqref{deltaS4fi} holds true even for matrix-valued functions $f_\alpha$. For this reason, Eq.\ \eqref{4bracketrelation} provides a consistent current algebra in (3+1)d even in the non-Abelian case.

\subsection{Generalization to arbitrary even spacetime dimensions}
\label{sec:higherdim} 

We present generalizations of the results in Section~\ref{sec:nbrackets} to arbitrary $d$-brackets where $d=2n$ is even. 

We start with the unordered operators $\hat{Q}_0(A_\alpha)$. 
It is straightforward to generalize our arguments for $d=2n=4$ to arbitrary $d=2n$ and thus show that $[\hat{Q}_0(A_1),\hat{Q}_0(A_2),\ldots,\hat{Q}_0(A_{d})]$ is a linear combination of $k$-body terms for  $k=1,2,\ldots,n$, and by defining a modified $d$-bracket $[\hat{Q}_0(A_1),\hat{Q}_0(A_2),\ldots,\hat{Q}_0(A_{d})]_{mod}$ by dropping all $k$-body terms with $k>1$ (i.e., keeping only the 1-body terms) one obtains 
\begin{align} 
[\hat{Q}_0(A_1),\hat{Q}_0(A_2),\ldots,\hat{Q}_0(A_{d})]_{mod} = \hat{Q}_0([A_1,A_2,\ldots,A_{d}]). 
\end{align} 
Motivated by our results for $d=2n=4$, we make the following ansatz for the modified $d$-bracket for the normal ordered operators $\hat{Q}(A_\alpha)$, 
\begin{align} 
[\hat{Q}(A_1),\hat{Q}(A_2),\ldots,\hat{Q}(A_{d})]_{mod} :&= \hat{Q}_0([A_1,A_2,\ldots,A_{d}])-R_{d}(A_1,A_2,\ldots,A_{d}) 
\end{align} 
with a multilinear and antisymmetric function $R_{d}$ to be found and 
\begin{align} 
 \hat{Q}_0([A_1,A_2,\ldots,A_{d}]) =  \hat{Q}([A_1,A_2,\ldots,A_{d}]) + b_{d}([A_1,A_2,\ldots,A_{d}]) 
\end{align} 
where  
\begin{equation} 
  b_{d}([A_1,A_2,\ldots,A_{d}]) =
  \mathrm{Tr} \big([A_1,A_2,\ldots,A_{d}]_{--}\big).
\end{equation} 
Thus, the key computation is the generalization of Eq.\ \eqref{b4computation} from $d=2n=4$ to arbitrary $d=2n$, 
\begin{align} 
  b_{d}([A_1,A_2,\ldots,A_{d}]) = \sum_{\epsilon_1,\epsilon_2,\ldots,\epsilon_{d-1}=\pm}
  \epsilon^{\alpha_1\alpha_2\ldots \alpha_{d}}
  \mathrm{Tr}\big(
  (A_{\alpha_1})_{-\epsilon_1} (A_{\alpha_2})_{\epsilon_1\epsilon_2}\cdots (A_{\alpha_{d}})_{\epsilon_{d-1}-} \big), 
\label{b2n} 
\end{align} 
etc. 
There is a single term in this sum which is finite for the pertinent one-particle operators in $d=2n$ spacetime dimensions, namely the term with alternating signs $\epsilon_j=(-1)^{j-1}$ for 
$j=1,\ldots,d-1$
\cite{Langmann1995},  
\begin{align} 
S_{d}(A_1,A_2,\ldots,A_{d}) &: = \epsilon^{\alpha_1\ldots\alpha_{d}} 
\mathrm{Tr} \big((A_{\alpha_1})_{-+}(A_{\alpha_2})_{+-}\cdots (A_{\alpha_{d}})_{+-}\big). 
\label{S2ndef} 
\end{align}  
Thus, we set $R_{d}$ equal to the remaining terms, 
\begin{align} 
R_{d}(A_1,A_2,\ldots,A_{d}) :=
b_{d}([A_1,A_2,\ldots,A_{d}])-S_{d}(A_1,A_2,\ldots,A_{d}). 
\end{align} 
Many of these remaining terms cancel by the cyclicity of the trace and the antisymmetry of the epsilon symbol, similarly as in the special case $d=2n=4$. 
It would be interesting to compute a simple formula for $R_{d}$ taking into account these cancellations and generalizing the formula for $R_4$ in Eq.\ \eqref{R4S4}, but this is left to future work. 

Thus, the result is 
\begin{align} 
[\hat{Q}(A_1), \ldots,\hat{Q}(A_{d})]_{mod} = \hat{Q}([A_1,\ldots,A_{d}]) + S_{d}(A_1,\ldots,A_{d}) 
\label{2nbracketrelation} 
\end{align} 
with 
\begin{align} 
  S_{d}(A_1,\ldots,A_{d})
  & = \frac{(-1)^{d/2-1}}{2^{d+1}}\epsilon^{\alpha_1\ldots\alpha_{d}}
    \mathrm{Tr}\big(F[F,A_{\alpha_1}][F,A_{\alpha_2}]\cdots [F,A_{\alpha_{d}}]\big)
    \nonumber \\
  & = \frac{(-1)^{d/2-1}}{2^{d}}\epsilon^{\alpha_1\ldots\alpha_{d}}
    \mathrm{Tr}_c\, \big(A_{\alpha_1} [F,A_{\alpha_2}]\cdots [F,A_{\alpha_{d}}]\big) 
\label{S2n}
\end{align} 
obtained from Eq.\ \eqref{S2ndef} using the notation introduced in Section \eqref{sec:NCG}; the latter formula makes precise that $S_{d}$ is a generalization to NCG of an integral of the de Rham form in $d-1$ dimensions. 
In particular, specializing the operators $A_\alpha$ to real-valued functions $f_\alpha$ on $M_{d-1}$ and $F$ to the sign of the Weyl operators in $d-1$ space dimensions (see the paragraph after Eq.\ \eqref{Trc2nd} for a more detailed description), identifying $\hat{Q}(A_\alpha)$ with $\hat{\varrho}(f_\alpha)$, and using the identity in Eq.\ \eqref{Trc2nd}, we obtain the following natural generalization of Eq.\ \eqref{4currentalgebra}, 
\begin{align} 
[\hat{\varrho}(f_1), \cdots,\hat{\varrho}(f_{d})]_{mod} 
= -
\left(
d/2\right)!\left(\frac{\ii}{2\pi}\right)^{d/2}  \int_{M_{d-1}} f_{1}\dd{f}_{2}\cdots \dd{f}_{d}, 
\label{2ncurrentalgebra} 
\end{align} 
using 
$
\int_{M_{d-1}} 
\epsilon^{\alpha_1\ldots\alpha_{d}} f_{\alpha_1}\dd{f}_{\alpha_2}\cdots \dd{f}_{\alpha_{d}}= d!\int_{M_{d-1}} f_{1}\dd{f}_{2}\cdots \dd{f}_{d}. 
$
It is known that the $d$-bracket satisfies the following generalized Jacobi identity
\cite{de_Azc_rraga_1997}, 
\begin{equation}
\frac1{d!(d-1)!}\epsilon^{\alpha_1\alpha_2\ldots\alpha_{2d-1}}[[A_{\alpha_1},\cdots,A_{d}],A_{d+1},\cdots,A_{2d-1}] =0
\end{equation} 
and, similarly as for $d=2n=4$, this suggests that the following consistency conditions should be fulfilled, $\delta S_{d}=0$ with 
\begin{equation} 
  \delta S_{d}(A_1,\ldots,A_{2d-1}) 
  := \frac1{d!(d-1)!}
  \epsilon^{\alpha_1\alpha_2\ldots\alpha_{2d-1}}
  S_{d}([A_{\alpha_1},\ldots,A_{d}],A_{d+1},\ldots,A_{2d-1}). 
\end{equation} 
In the special case $A_\alpha=f_\alpha$ etc.\ leading to Eq.\ \eqref{2ncurrentalgebra}, 
$S_{d}\propto \int_{M_{d-1}}f_{1}\dd{f}_{2}\cdots \dd{f}_{d}$ and $[f_1,\ldots,f_{d}]=0$ imply 
\begin{equation} 
\delta S_{d}(f_1,\cdots,f_{2d-1})=0, 
\end{equation} 
which proves that our $d$-dimensional current algebra in Eq.\ \eqref{2ncurrentalgebra} is consistent. It would be interesting to generalize Eq.\ \eqref{deltaS4} from $d=4$ to general $d={\it even}$, but this is left to future work.

\subsection{Non-Abelian current algebras in even spacetime dimensions}\label{sec:nonAbelian} 
In this paper, we emphasize the Abelian case where the operators $A_\alpha$ are real-valued functions $f_\alpha$ which commute, $[f_\alpha,f_\beta]=0$; this corresponds to the important special case of $U(1)$ current algebras. However, it is also interesting to consider matrix-valued functions $f_\alpha$ corresponding to a theory where the fermions have an additional color index. 
As we now show, our general results allow for such non-Abelian current algebras in higher dimensions.

Consider a theory of Weyl fermions on $(d-1)=(2n-1)$-dimensional space $M_{d-1}$ (which either is $\mathbb{R}^{d-1}$ or a $(d-1)$-dimensional torus), 
\begin{align}
  \hat{H}_{{\it Weyl}} =\int_{M_{d-1}} \dd^{d-1}\mathbf{x}\, \sum_{i=1}^{d-1} 
  \normord{\hat{\psi}^{\dag}_a (-\ii \Gamma_{i}\partial_{i} ) \hat{\psi}_a},
\label{d-1 dim Weyl}
\end{align}
where 
$\{\Gamma_i\}_{i=1,\cdots, d-1}$
are hermitian $\nu\times\nu$-matrices, $\nu=2^{d/2-1}$, satisfying the relations $\{\Gamma_i,\Gamma_j\}=2\delta_{ij}$, and the fermion operators carry a color index $a=1,\ldots,N$ (in addition to the index corresponding to the $\Gamma_i$-matrices which we suppress) . For example, the case $N=2$ corresponds to the case of fermions where the color index can be identified with spin $a=\uparrow,\downarrow$ which, 
in addition to charge transport, can also describe the transport of spin.  
In such a theory, we can consider the non-Abelian current operators
\begin{align} 
\varrho(f) := \int_{M_{d-1}} 
\dd^{d-1}\mathbf{x}\,  f_{ab}(\mathbf{x}) \normord{\hat{\psi}^{\dag}_a(\mathbf{x})\hat{\psi}^{\phantom\dag}_b(\mathbf{x})}
\end{align} 
where $f(\mathbf{x})=\{ f_{ab}(\mathbf{x})\}_{a,b=1}^N$ are functions on space $M_{d-1}$ with values in the hermitian $N\times N$-matrices. We then can get the $d$-brackets of these currents from our general result in Eqs.\ \eqref{2nbracketrelation}--\eqref{S2n} by choosing as Hilbert space the space of square-integrable functions on 
$M_{d-1}$ with values in vectors $\mathbb{C}_{spin}^{\nu}\otimes\mathbb{C}^{N}_{color}$, $F$ the sign of the 1-particle Weyl operator $\sum_{i=1}^{d-1} (-\ii \Gamma_{i}\partial_{i})$, and $A_\alpha=f_\alpha$ (matrix-valued functions on $M_{d-1}$). Using the following known non-Abelian generalization of Eq.\ \eqref{Trc2nd}
\cite{Langmann1995}, 
\begin{align} 
\mathrm{Tr}_c(f_1[F,f_2]\cdots [F,f_{d}]) =  \left(-\frac{\ii}{2\pi}\right)^{d/2} \frac{2^{d} (d/2)!}{d!}\int_{M_{d-1}} \mathrm{tr}_N(f_1\dd{f}_2\cdots \dd{f}_{d}) 
\label{Trc2ndgen}
\end{align} 
where $\mathrm{tr}_N$ is the usual trace of $N\times N$ matrices, this yields
\begin{align} 
[\hat{\varrho}(f_1),
\hat{\varrho}(f_2),\ldots,
\hat{\varrho}(f_{d})]_{mod} = \hat{\varrho}([f_1,f_2,\ldots,f_{d}])-\left(\frac{\ii}{2\pi}\right)^{d/2}\frac{(d/2)!}{d!} \int_{M_{d-1}} \epsilon^{\alpha_1\alpha_2\ldots,\alpha_{d}} \mathrm{tr}_N(f_{\alpha_1} \dd{f}_{\alpha_2} \cdots \dd{f}_{\alpha_{d}}) . 
\label{nonAbelian}
\end{align} 
Note that, for $d=2n=2$, we recover the well-known non-Abelian current algebra in (1+1)d (see e.g.\ \cite{Carey1987}] 
$$
[\hat{\varrho}(f_1),
\hat{\varrho}(f_2)] = 
\hat{\varrho}([f_1,f_2])-\frac{\ii}{4\pi}\int_{M_1}\epsilon^{\alpha\beta}\mathrm{tr}_N(f_\alpha \dd{f}_\beta)
$$
which is consistent (as discussed in Section~\ref{sec:2ndquantizationmap}). For $2n=4$, we obtain 
\begin{equation} 
[\hat{\varrho}(f_1),
\hat{\varrho}(f_2),
\hat{\varrho}(f_3),
\hat{\varrho}(f_4)]_{mod} = \hat{\varrho}([f_1,f_2,f_3,f_4])-\frac{1}{48\pi^2}\int_{M_3}
\epsilon^{\alpha\beta\gamma\delta}  \mathrm{tr}_N(f_\alpha \dd{f}_\beta \dd{f}_\gamma \dd{f}_\delta). 
\label{nonAbelian4} 
\end{equation} 
As shown in the following paragraph, 
Eq.\ \eqref{deltaS4} and known results in the literature
\cite{Langmann_1994} imply that this non-Abelian current algebra in (3+1)d is consistent as well. For $d=2n>4$, we do not have a proof that Eq.\ \eqref{nonAbelian} is consistent (it certainly would be interesting to find such a proof).

To conclude this section, we prove the consistency of Eq.\ \eqref{nonAbelian4}, i.e., we show that $\delta S_4(f_1,\ldots,f_7)=0$   holds true even if the function $f_\alpha$ are matrix-valued. For that, we first compute $\delta S_4(f_1,\ldots,f_7)$ by specializing Eq.\ \eqref{deltaS4}  to $A_\alpha=f_\alpha$ and computing the resulting expression in a gradient expansion, using symbol calculus of pseudodifferential operators
\cite{Langmann_1994}.
We use that the symbol of the operator $F$ is $F(\mathbf{p})=\sum_{i=1}^3\sigma_i p_i/|\mathbf{p}|$ (where $p_i$ are the components of the momentum $\mathbf{p})$, the symbol of the operator $[F,f_\alpha]$ is 
$$
(-\ii)\sum_{i=1}^3 \frac{\partial F(\mathbf{p})}{\partial p_i}\frac{\partial f_\alpha(\mathbf{x})}{\partial x_i }; 
$$
this implies that the leading term in the gradient expansion of $\delta S_4(f_1,\ldots,f_7)$ obtained from Eq.\ \eqref{deltaS4} is a linear combination of terms 
$$
\int_{M_3}\epsilon^{\alpha_1\ldots\alpha_7}\mathrm{tr}_N\Bigl(f_{\alpha_1}f_{\alpha_2}\partial_{i_1} f_{\alpha_3}[f_{\alpha_4},\partial_{i_2} f_{\alpha_5}\partial_{i_3} f_{\alpha_6}] \partial_{i_4} f_{\alpha_7} \Bigr) 
$$ 
where $\partial_i f_\alpha$ is short for $\frac{\partial f_\alpha}{\partial x_i}$ (see Ref.\ \cite{Langmann_1994} for details). Thus, Eq.\ \eqref{deltaS4} implies that $\delta S_4(f_1,\ldots,f_4)$ is an integral of terms with at least 4 differentiations. On the other hand, Eqs.\ \eqref{S4} and \eqref{Trc2ndgen} imply 
$$
S_4(f_1,f_2,f_3,f_4)= -\frac{1}{48\pi^2}\int_{M_3}
\epsilon^{\alpha\beta\gamma\delta}  \mathrm{tr}_N(f_\alpha \dd{f}_\beta \dd{f}_\gamma \dd{f}_\delta), 
$$
and computing $\delta S_4(f_1,\ldots,f_7)= (1/144)\epsilon^{\alpha_1\ldots\alpha_7}S_4([[f_{\alpha_1},f_{\alpha_2},f_{\alpha_3},f_{\alpha_4}],f_{\alpha_5},f_{\alpha_6},f_{\alpha_7}])$
from this one finds that it is an integral of terms with 3 differentiations. This leads to a contradiction unless $\delta S_4(f_1,\ldots,f_7)=0$,
which proves the result.

\section{Bulk density operator algebra}
\label{Bulk density operator algebra}
In this section,
we discuss the bulk density operator algebra for (topological) band insulators.
We first warm up by discussing (2+1)d Chern insulators, and then look at
higher-dimensional topological insulators.

\subsection{Set up}
\label{Setup}

Let us start by introducing the necessary notations for one-particle lattice Hamiltonians defined on a $d$-dimensional hypercubic lattice whose lattice constant is $\mathfrak{a}$.
Consider a tight-binding Hamiltonian,
\begin{align}
\hat H
=
\sum_{\bx,\bx'}
\hat{c}^{\dag}(\bx)\,
\mathcal{H}(\bx,\bx')\,
\hat{c}(\bx'),
\label{tight binding}
\end{align}
where $\hat{c}(\bx)$ is an $N_f$-component fermion annihilation operator, and index $\bx = (x_1,\dots,x_d) \in (\mathfrak{a}\mathbb{Z})^d = \{\mathfrak{a} \bm{n} | \bm{n} \in \Z^d\}$ labels a site on a $d$-dimensional lattice(the internal indices are suppressed).
The volume of the unit cell is $\Omega = \mathfrak{a}^d$. 
Each block in the single particle Hamiltonian $\mathcal{H}(\bx,\bx')$ is an $N_f\times N_f$ matrix, and subjected to the hermiticity condition $\mathcal{H}(\bx',\bx)^\dag =\mathcal{H}(\bx,\bx')$.
The components in $\hat{c}(\bx)$ can describe, e.g., orbitals or spin degrees of freedom, as well as different sites within a crystal unit cell centered at $\bx$.
\footnote{
In this paper, for simplicity, we assume that all internal degrees of freedom 
are spatially localized at the center of the unit cell, $\bx$.
In other words, we set the Fourier transformation to momentum space as
$\ket{\bk,i} = \sum_{\bx \in (\mathfrak{a}\Z)^d} e^{\ii k \cdot \bm{x}} \ket{\bk,i}$,
where $i$ represents the internal degrees of freedom in the unit cell,
and $\ket{\bk,i}$ is periodic in the BZ.
Within this treatment, 
the Bloch function $\ket{\phi^a(\bk)}$ has no extra phases depending on the
localized position in the unit cell, i.e.,
$\braket{\bk,i|\phi^a(\bk)} = e^{\ii \bk \cdot \bx} {u_i}^a(\bk)$.
This simplification can not be applied to effects
that depend on the spatial embedding of the internal degrees 
of freedom, for instance, the semiclassical equations of motion. }

Provided the system has translational symmetry, $\mathcal{H}(\bx,\bx') =
\mathcal{H}(\bx-\bx')$, with periodic boundary conditions in each spatial direction (i.e., the system is defined on a torus $T^d$).
Hereafter, we take a thermodynamic limit.
Performing the Fourier transformation, we obtain the Hamiltonian in momentum space,
\begin{align}
\hat{H}
=
\Omega \int_{T^d} \frac{\dd^d \boldsymbol{k}}{(2 \pi)^d}\,
\hat{c}^{\dag}(\boldsymbol{k}) \,
\mathcal{H}(\boldsymbol{k})\,
\hat{c} (\boldsymbol{k}) ,
\end{align}
where the crystal momentum $\boldsymbol{k}$ runs over the first Brillouin zone torus $T^d = \left[-{\pi}/{\mathfrak{a}},
{\pi}/{\mathfrak{a}}\right]^d$, and the Fourier component of the fermion operator and the Hamiltonian are given by
\begin{align}
\hat{c}(\boldsymbol{x})
&=
\Omega \int_{T^d} \frac{\dd^d \boldsymbol{k}}{(2 \pi)^d}
e^{\ii \boldsymbol{k} \cdot \boldsymbol{x}}
\hat{c}(\boldsymbol{k}),
\nonumber\\
\mathcal{H}(\boldsymbol{k})
&=
\sum_{\boldsymbol{x} \in (\mathfrak{a}\mathbb{Z})^d}
e^{- \ii \boldsymbol{k} \cdot \boldsymbol{x}}
\mathcal{H}(\bm {x}).
\end{align}

The Bloch Hamiltonian $\mathcal{H}(\boldsymbol{k})$ is diagonalized by $N_f$ vectors 
\begin{align}
\mathcal{H}(\bk) u^a(\bk)
=
\epsilon^{a}(\bk) u^a(\bk),
\quad 
{a} = 1,\ldots, N_{f}, 
\end{align}
where the eigenvectors are normalized as $[u^a(\bk)]^\dag u^b(\bk)=\sum_{i=1}^{N_f} [u^a_i(\bk)]^* u^b_i(\bk) = \delta^{ab}$. 
The fermion field operator can be expanded in terms of the eigenvectors as
\begin{align}
\hat{c}_i(\bk)
=
\sum_{a=1}^{N_f} u^a_i(\bk) \hat{\chi}_a(\bk),
\quad
i = 1,\ldots, N_{f},
\end{align}
where
$\hat{\chi}_{{a}}(\boldsymbol{k})$ represents
a fermionic operator in the eigenbasis
and given by
\begin{align}
\hat{\chi}_{a}(\boldsymbol{k})
=
\sum_{i=1}^{N_f} \big[u_{i}^{a}(\boldsymbol{k})\big]^*
\hat{c}_i (\boldsymbol{k}).
\end{align}

Below, we will focus on a particular set of bands;
we focus on $N_-$ bands below the Fermi energy and we assume they are separated by the other bands by an energy gap.
We then project out $N_+$ bands for each $\boldsymbol{k}$ with $N_{+} + N_{-} = N_{f}$.
We call the set of filled Bloch eigenbases $\{u^{\hat a}(\bk)\}$, where hatted indices $\hat{a}=1,\ldots, N_-$ labels the bands of our interest only.
From here on, we will use the shorthand notation $\braket{u^a(\bk)|u^b(\bk')} = \sum_{i=1}^{N_f} [u_i^a(\bk)]^* [u_i^b(\bk')]$.

\subsection{The bulk algebra in (2+1)d}
In this section, we summarize a commutator relation of the projected density operator for Chern insulators in the bulk
\cite{Parameswaran2012, Murthy2012}.
(2+1)d Chern insulators are those insulators characterized by non-zero first Chern number
in momentum space:
\begin{align}
Ch_1 = \frac{\ii}{2 \pi} \int \dd^2\boldsymbol{k}\,
\sum_{\hat{a}=1}^{N_-}
\mathcal{F}^{\hat{a}\hat{a}}_{xy}(\boldsymbol{k}),
\end{align}
where $\mathcal{F}$ is the Berry curvature
and given in terms of the Berry connection
$\mathcal{A}$
as 
\begin{align}
\mathcal{F}^{\hat a\hat b} &= [\dd \mathcal{A} + \mathcal{A} \wedge \mathcal{A}]^{\hat a\hat b} = \dd \mathcal{A}^{\hat a\hat b} + \sum_{\hat c=1}^{N_-} \mathcal{A}^{\hat a\hat c} \wedge \mathcal{A}^{\hat c\hat b},
\nonumber \\
\mathcal{A}^{\hat a\hat b} &= \braket{u^{\hat a}|\dd u^{\hat b}}. 
\end{align}

Because of the presence of the bulk energy gap, 
focusing on low energies,
we consider dynamics within the occupied bands.
The projected fermion field operator $\hat \psi_i(\bx)$ is given by restricting the band sum to occupied states,
\begin{align}
&\hat{\psi}_i(\bk) := \sum_{\hat a=1}^{N_-} u_i^{\hat a}(\bk) \hat{\chi}_{\hat a}(\bk), \label{eq:projected_fermion_op_k}\\
&\hat{\psi}_i(\bx) = \Omega \int_{T^d} \frac{\dd^d\boldsymbol{k}}{(2\pi)^d} e^{\ii \bk \cdot \bx} \hat{\psi}_i(\bk),
\label{psi expansion}
\end{align}
and the projected density operator is defined by
\begin{align}
\hat{\rho}(\bx) := \sum_{i=1}^{N_f} \hat \psi^{\dag}_i(\bx) \hat \psi_i(\bx).
\end{align}
Note that in Eq.\ \eqref{eq:projected_fermion_op_k}, the eigenvectors $u^{\hat
  a}(\bk)$ and the fermion operators $\hat \chi_{\hat a}(\bk)$ might not be
globally defined over the BZ for each. These are separately gauge-dependent
under the gauge transformation $u^{\hat a}(\bk) \mapsto \sum_{\hat b=1}^{N_-}
u^{\hat b}(\bk) [V(\bk)]_{\hat b \hat a}$ and $\hat \chi^{\hat a}(\bk) \mapsto
\sum_{\hat b=1}^{N_-} [V(\bk)]^*_{\hat b \hat a} \chi^{\hat b}(\bk)$,
where $V(\bk) \in U(N_-)$. However, the projected fermion operator $\hat{\psi}_i(\bk)$ remains gauge-independent.

The Fourier transformation of $\hat{\rho}(\bx)$ is given by
$\hat{\rho}(\bx) = 
\Omega \int \frac{\dd^d q}{(2 \pi)^d} 
e^{\ii \boldsymbol{q} \cdot \bx} \hat{\rho}(\boldsymbol{q})$,
and $\hat{\rho}(\boldsymbol{q})$ 
is expanded in terms of the Bloch basis 
$\hat{\chi}_{\hat a}(\boldsymbol{k})$ as
\begin{align}
\hat{\rho}(\boldsymbol{q}) &=
\Omega \int_{T^d} \frac{\dd^d \boldsymbol{k}}{(2\pi)^d} \sum_{\hat a,\hat b=1}^{N_-}
\braket{u^{\hat a}(\boldsymbol{k})|u^{\hat b}(\boldsymbol{k}+\boldsymbol{q})}
\hat{\chi}^{\dag}_{\hat a}(\boldsymbol{k})
\hat{\chi}_{\hat b}(\boldsymbol{k}+\boldsymbol{q}).
\label{rho expansion}
\end{align}
The matrix elements of $\hat{\rho}(\boldsymbol{q})$ with respect to the
one-particle basis $\{ \hat{\chi}^\dag_{\hat a}(\bk)\ket{0}\}_{\bk \in T^d, \hat
  a =1,\dots,N_-}$
(which are $U(N_-)$ gauge-dependent) are
\begin{align}
[\hat{\rho}(\boldsymbol{q})]_{\boldsymbol{k} \hat a;\boldsymbol{k}' \hat b}
&= \frac{(2\pi)^d}{\Omega} \delta^{d}(\boldsymbol{k}'-\boldsymbol{k}-\boldsymbol{q})
\braket{u^{\hat a}(\boldsymbol{k})|u^{\hat b}(\boldsymbol{k}+\boldsymbol{q})}.
\end{align}
A product of two $[\hat{\rho}(\boldsymbol{q})]$ has a simple form
\begin{align}
([\hat{\rho}(\boldsymbol{q})]
[\hat{\rho}(\boldsymbol{q}')])_{\boldsymbol{k} \hat a;\boldsymbol{k}' \hat b}
&= \frac{(2\pi)^d}{\Omega} \delta^{d}(\boldsymbol{k}'-\boldsymbol{k}-\boldsymbol{q}-\boldsymbol{q}')
\nonumber \\
&
\quad
\times
\sum_{\hat c=1}^{N_-} \braket{u^{\hat a}(\boldsymbol{k})|u^{\hat c}(\boldsymbol{k}+\boldsymbol{q})}\braket{u^{\hat c}(\boldsymbol{k}+\boldsymbol{q})|u^{\hat b}(\boldsymbol{k}+\boldsymbol{q}+\boldsymbol{q}')}.
\label{ProdRhoChi}
\end{align}
There is a similar expression for $[\hat{\rho}(\boldsymbol{q}_1)][\hat{\rho}(\boldsymbol{q}_2)]\cdots [\hat{\rho}(\boldsymbol{q}_n)]$ for any $n$.

Let us now evaluate the commutator of the projected density operators.
Naively,
we expect that it should have the same properties as the integer QHE in the long wavelength limit.
We note that, generically, for second quantized one-particle operators $\hat{Q}_0(A) = \sum_{mn} A_{mn} \hat{\psi}^{\dag}_m \hat{\psi}^{\ }_n$, their commutator is given by $[\hat{Q}_0(A),\hat{Q}_0(B)] = \hat{Q}_0([A,B])$.
Thus the commutator of $\hat{\rho}(\boldsymbol{q})$ is easily obtained from the product formula (\ref{ProdRhoChi}),
\begin{align}
[\hat{\rho}(\boldsymbol{q}_1),\hat{\rho}(\boldsymbol{q}_2)]
&= \sum_{\hat a, \hat b} \Omega \int_{T^d} \frac{\dd^{d}\boldsymbol{k}}{(2\pi)^{d}} \sum_{\hat c=1}^{N_-}
\epsilon^{i_1 i_2}
\braket{u^{\hat a}(\boldsymbol{k})|u^{\hat c}(\boldsymbol{k}+\boldsymbol{q}_{i_1})}
\nonumber \\
&\quad
\times
\braket{u^{\hat c}(\boldsymbol{k}+\boldsymbol{q}_{i_1})|u^{\hat b}(\boldsymbol{k}+\boldsymbol{q}_1+\boldsymbol{q}_2)}
  \hat{\chi}^{\dag}_{\hat a}(\boldsymbol{k})
  \hat{\chi}_{\hat b}(\boldsymbol{k}+ \boldsymbol{q}_1+\boldsymbol{q}_2).
\end{align}
In the long-wavelength limit,
we can show the following relation
\begin{align}
&
\sum_{\hat c=1}^{N_-} \epsilon^{i_1 i_2} \braket{u^{\hat a}(\boldsymbol{k})|u^{\hat c}(\boldsymbol{k}+\boldsymbol{q}_{i_1})}
  \braket{u^{\hat c}(\boldsymbol{k}+\boldsymbol{q}_{i_1})|
  u^{\hat b}(\boldsymbol{k}+\boldsymbol{q}_{i_1}+\boldsymbol{q}_{i_2})}
\nonumber \\
&\quad
=
(\boldsymbol{q}_1 \wedge \boldsymbol{q}_2) \epsilon^{\mu_1 \mu_2} \frac{\mathcal{F}^{\hat a \hat b}_{\mu_1 \mu_2}(\boldsymbol{k})}{2} + O(q^3),
\label{wfn overlaps}
\end{align}
where $\boldsymbol{q}_1 \wedge \boldsymbol{q}_2 = \epsilon^{\mu_1 \mu_2} (q_1)_{\mu_1} (q_2)_{\mu_2} = (q_1)_x (q_2)_y - (q_1)_y (q_2)_x$,
and $\mathcal{F}_{\mu\nu}(\boldsymbol{k})$
is the $N_- \times N_-$ non-Abelian Berry curvature.
Thus, we obtain
\begin{align}
[\hat{\rho}(\boldsymbol{q}_1),\hat{\rho}(\boldsymbol{q}_2)]
&= (\boldsymbol{q}_1 \wedge \boldsymbol{q}_2) \sum_{\hat a, \hat b} \Omega \int_{T^d} \frac{\dd^{d}\boldsymbol{k}}{(2\pi)^{2}}
\epsilon^{\mu_1 \mu_2} \frac{\mathcal{F}^{\hat a \hat b}_{\mu_1 \mu_2}(\boldsymbol{k})}{2}
\hat{\chi}^{\dag}_{\hat a}(\boldsymbol{k}) 
\hat{\chi}_{\hat b}(\boldsymbol{k}+\boldsymbol{q}_1+\boldsymbol{q}_2)
+ O(q^3).
\end{align}

If the Berry curvature were to be constant over the BZ,
$\mathcal{F}(\boldsymbol{k}) = {const.}$,
the projected density operator would obey the $w_{\infty}$ algebra in the long wavelength limit,
\begin{align}
[\hat{\rho}(\boldsymbol{q}_1),\hat{\rho}(\boldsymbol{q}_2)]
&= -\ii  \Omega \frac{Ch_1}{2 \pi} (\boldsymbol{q}_1 \wedge \boldsymbol{q}_2) \hat{\rho}(\boldsymbol{q}_1+\boldsymbol{q}_2) + O(q^3)
\qquad
(\mathrm{for} \ \mathcal{F}(\boldsymbol{k}) = {const.}),
\end{align}
as in the integer QHE.
Here, $Ch_1$ is the first Chern number
of the bulk Chern insulator.
Generically, however, the Berry curvature
is not constant in the BZ;
The density operator algebra in generic Chern insulators
is given by neither
$W_{1+\infty}$ or $w_{\infty}$ algebra, and not directly related
to the first Chern number $Ch_1$.
Nevertheless,
as we will show below,
the topological invariant $Ch_1$ still shows up in the current algebra
if one considers a many-body ground state in the presence of a boundary.

\subsection{The bulk algebra in generic even space dimensions}

We now generalize  
the (2+1)d results presented in the previous section 
to higher dimensional topological insulators.
We start with the product of $d$ projected density operators
for $(d+1)$-dimensional topological insulators.
It is given in terms of 
the $d$-bracket of the corresponding single-particle
operators as
\begin{align}
&
\big[
\hat{\rho}(\boldsymbol{q}_1),\,
\cdots,
\hat{\rho}(\boldsymbol{q}_d)
\big]_{{mod}}
= \hat{Q}_0\big(\big[[\hat{\rho}(\boldsymbol{q}_1)],
\cdots,
[\hat{\rho}(\boldsymbol{q}_d)]\big]\big),
\label{mod four-bracket four density}
\end{align}
where $[\hat{\rho}(\boldsymbol{q})]$ 
on the RHS represents
a single-particle operator of $\hat{\rho}(\boldsymbol{q})$ with respect to the single-particle basis $\{ \hat{\chi}^\dag_{\hat a}(\bk)\ket{0}\}_{\bk \in T^d, \hat a =1,\dots,N_-}$. 
We will show that the RHS in Eq.\ \eqref{mod four-bracket four density} is evaluated as
\begin{align}
  &
\hat{Q}_0\big(\big[[\hat{\rho}(\boldsymbol{q}_1)],\dots,[\hat{\rho}(\boldsymbol{q}_d)]\big]\big)
\nonumber \\
&\quad = \sum_{\hat a, \hat b=1}^{N_-} \Omega \int_{T^d} \frac{\dd^{d}\boldsymbol{k}}{(2\pi)^{d}}
(\boldsymbol{q}_1 \wedge \cdots \wedge \boldsymbol{q}_d) \epsilon^{\mu_1 \dots \mu_d}
\nonumber \\
&\qquad
\times
\sum_{\hat c_1, \dots, \hat c_{d-1} = 1}^{N_-}
\frac{\mathcal{F}^{\hat a \hat c_1}_{\mu_1\mu_2}(\bk)}{2} \cdots \frac{\mathcal{F}^{\hat c_{d-1} \hat b}_{\mu_{d-1}\mu_d}(\bk)}{2}
\hat{\chi}^{\dag}_{\hat a}(\boldsymbol{k}) 
\hat{\chi}_{\hat b}(\boldsymbol{k}+\boldsymbol{q}_1+\cdots +\boldsymbol{q}_d) + O(q^{d+1})
\label{main result}
\end{align}
for even $d$,
where we denoted 
$(\boldsymbol{q}_1 \wedge \cdots \wedge \boldsymbol{q}_d)=
\mathrm{det}\, (
\boldsymbol{q}_1, \cdots, \boldsymbol{q}_d)
= \epsilon^{\mu_1 \cdots \mu_d} (q_1)_{\mu_1} \cdots (q_d)_{\mu_d}$.
To show Eq.\ \eqref{main result},
we start by noting that  
the Berry curvature $\mathcal{F}$ is related with the projection matrix 
$P = \sum_{\hat a=1}^{N_-} \ket{u^{\hat a}}\bra{u^{\hat a}}$ by
\begin{align}
\mathcal{F}^{\hat a\hat b}
= \Braket{\dd u^{\hat a}|(1-P)|\dd u^{\hat b}}
= \Braket{u^{\hat a}|\dd P \dd P|u^{\hat b}}.
\end{align}
Furthermore, the projected density operator is given in terms of the projection matrix as
\begin{align}
\hat{\rho}(\boldsymbol{q})
&= \Omega \int_{T^d} \frac{\dd^{d}\boldsymbol{k}}{(2\pi)^{d}}
\sum_{\hat{a},\hat{b}}
\hat{\chi}^{\dag}_{\hat a}(\boldsymbol{k}) 
\braket{u^{\hat a}(\boldsymbol{k})|u^{\hat b}(\boldsymbol{k}+\boldsymbol{q})} 
\hat{\chi}_{\hat b}(\boldsymbol{k}+\boldsymbol{q})
\nonumber \\
&= \Omega \int_{T^d} \frac{\dd^{d}\boldsymbol{k}}{(2\pi)^{d}}
\sum_{i,j}
\hat{c}^{\dag}_{i}(\boldsymbol{k}) [P(\boldsymbol{k})P(\boldsymbol{k}+\boldsymbol{q})]_{ij} 
\hat{c}_{j}(\boldsymbol{k}+\boldsymbol{q}).
\end{align}
We will find the following relations useful :
\begin{align}
& P^2 = P,
\quad
P \partial_{\mu} P = \partial_{\mu} P (1-P),
\quad
P \partial_{\mu} P P = 0 ,
\nonumber \\
&
P \partial_{\mu} P \partial_{\nu} P = \partial_{\mu} P \partial_{\nu} P P,
\quad
\partial_{\mu} \partial_{\nu} P P = - P \partial_{\mu} P \partial_{\nu}P -P \partial_{\nu}P \partial_{\mu} P.
\end{align}
To calculate the four-bracket 
$\big[[\hat{\rho}(\boldsymbol{q}_1)],\dots,[\hat{\rho}(\boldsymbol{q}_d)]\big]
$, 
we consider 
the following antisymmetrized product 
\begin{align}
(*)=
\epsilon^{i_1i_2i_3i_4\dots } P(\boldsymbol{k}) P(\boldsymbol{k}+\boldsymbol{q}_{i_1})
P(\boldsymbol{k}+\boldsymbol{q}_{i_1}+\boldsymbol{q}_{i_2}) P(\boldsymbol{k}+\boldsymbol{q}_{i_1}+\boldsymbol{q}_{i_2}+\boldsymbol{q}_{i_3})
P(\boldsymbol{k}+\boldsymbol{q}_{i_1}+\boldsymbol{q}_{i_2}+\boldsymbol{q}_{i_3}+\boldsymbol{q}_{i_4}) \cdots
\end{align}
By noting
\begin{align}
&\frac{1}{2}P(\boldsymbol{k})[P(\boldsymbol{k}+\boldsymbol{q})-P(\boldsymbol{k}+\boldsymbol{q}')] P(\boldsymbol{k}+\boldsymbol{q}+\boldsymbol{q}')
\nonumber \\
&= \frac{1}{2}P \Big[ (q_{\mu}-q'_{\mu})\partial_{\mu}P + \frac{1}{2} (q_{\mu} q_{\nu} - q'_{\mu} q'_{\nu}) \partial_{\mu}\partial_{\nu}P \Big]
\Big[ P + (q_{\mu}+q'_{\mu}) \partial_{\mu}P \Big] + O(q^3)
\nonumber \\
&= \frac{1}{2}P \Big[ (q_{\mu}-q'_{\mu})(q_{\nu}+q'_{\nu}) \partial_{\mu}P \partial_{\nu}P + \frac{1}{2} (q_{\mu} q_{\nu}
- q'_{\mu} q'_{\nu}) \partial_{\mu}\partial_{\nu}P P \Big] + O(q^3)
\nonumber \\
&= \frac{1}{2}P
\Big[ (q_{\mu}-q'_{\mu})(q_{\nu}+q'_{\nu}) \partial_{\mu}P \partial_{\nu}P - \frac{1}{2} (q_{\mu} q_{\nu} - q'_{\mu} q'_{\nu}) ( P \partial_{\mu} P \partial_{\nu} P -P \partial_{\nu} P\partial_{\mu} P )
\Big] + O(q^3)
\nonumber \\
&= \frac{1}{2} (q_{\mu}q'_{\nu}-q_{\nu}q'_{\mu}) P \partial_{\mu}P \partial_{\nu}P + O(q^3),
\end{align}
then, we can show
\begin{align}
(*)
&= \epsilon^{i_1i_2\dots } \frac{1}{2} P(\boldsymbol{k}) [P(\boldsymbol{k}+\boldsymbol{q}_{i_1})-P(\boldsymbol{k}+\boldsymbol{q}_{i_2})]
P(\boldsymbol{k}+\boldsymbol{q}_{i_1}+\boldsymbol{q}_{i_2})
\nonumber \\
&\quad  \cdot
\frac{1}{2}P(\boldsymbol{k}+\boldsymbol{q}_{i_1}+\boldsymbol{q}_{i_2})
[ P(\boldsymbol{k}+\boldsymbol{q}_{i_1}+\boldsymbol{q}_{i_2}+\boldsymbol{q}_{i_3})
-P(\boldsymbol{k}+\boldsymbol{q}_{i_1}+\boldsymbol{q}_{i_2}+\boldsymbol{q}_{i_4})]
P(\boldsymbol{k}+\boldsymbol{q}_{i_1}+\boldsymbol{q}_{i_2}+\boldsymbol{q}_{i_3}+\boldsymbol{q}_{i_4}) \cdots
\nonumber \\
&= \epsilon^{i_1i_2\dots }
\Big[ \frac{1}{2} ((q_{i_1})_{\mu_1}(q_{i_2})_{\mu_2}-(q_{i_1})_{\mu_2}(q_{i_2})_{\mu_1}) P \partial_{\mu_1}P \partial_{\mu_2}P + O(q^3) \Big]
\nonumber \\
&\quad  \cdot \Big[ \frac{1}{2} ((q_{i_3})_{\mu_3}(q_{i_4})_{\mu_4}-(q_{i_3})_{\mu_4}(q_{i_4})_{\mu_3}) P \partial_{\mu_3}P \partial_{\mu_4}P + O(q^3) \Big] \cdots
\nonumber \\
&= \epsilon^{i_1i_2\dots }
\Big[ (q_{i_1})_{\mu_1}(q_{i_2})_{\mu_2} P \partial_{\mu_1}P \partial_{\mu_2}P + O(q^3) \Big]
\Big[ (q_{i_3})_{\mu_3}(q_{i_4})_{\mu_4} P \partial_{\mu_3}P \partial_{\mu_4}P + O(q^3) \Big] \cdots .
\end{align}
In particular, 
for the product of $d+1$ projectors 
and when the space dimension is $d$
(here $d$ is even), 
\begin{align}
(*)
= (\boldsymbol{q}_1 \wedge \cdots \wedge \boldsymbol{q}_d) \epsilon^{\mu_1 \dots \mu_d} P \partial_{\mu_1}P \cdots \partial_{\mu_d}P + O(q^{d+1}).
\label{product of Ps}
\end{align}
Finally, by noticing 
\begin{align}
&\Braket{u^{\hat a}(\boldsymbol{k})| 
P(\boldsymbol{k}) P(\boldsymbol{k}+\boldsymbol{q}_{i_1}) P(\boldsymbol{k}+\boldsymbol{q}_{i_1}+\boldsymbol{q}_{i_2})
\cdots P(\boldsymbol{k}+\boldsymbol{q}_{i_1}+\cdots +\boldsymbol{q}_{i_d}) |
u^{\hat b}(\boldsymbol{k}+\boldsymbol{q}_{i_1}+\cdots +\boldsymbol{q}_{i_d})}
\nonumber \\
&= \sum_{\hat c_1, \dots, \hat c_{d-1} = 1}^{N_-}
\Braket{u^{\hat a}(\boldsymbol{k})|u^{\hat c_1}(\boldsymbol{k}+\boldsymbol{q}_{i_1})}
\Braket{u^{\hat c_1}(\boldsymbol{k}+\boldsymbol{q}_{i_1})|u^{\hat c_2}(\boldsymbol{k}+\boldsymbol{q}_{i_1}+\boldsymbol{q}_{i_2})}
\nonumber \\
&
\qquad \qquad
\cdots
\times
\Braket{u^{\hat c_{d-1}}(\boldsymbol{k}+\boldsymbol{q}_{i_1}+\cdots +\boldsymbol{q}_{i_{d-1}})|u^{\hat b}(\boldsymbol{k}+\boldsymbol{q}_{i_1}+\cdots +\boldsymbol{q}_{i_{d}})},
\end{align}
Eq.\ \eqref{product of Ps} leads to the main result \eqref{main result}
since
\begin{align}
&\epsilon^{i_1\dots i_d}
\sum_{\hat c_1, \dots, \hat c_{d-1} = 1}^{N_-}
\Braket{u^{\hat a}(\boldsymbol{k})|u^{\hat c_1}(\boldsymbol{k}+\boldsymbol{q}_{i_1})}
\Braket{u^{\hat c_1}(\boldsymbol{k}+\boldsymbol{q}_{i_1})|u^{\hat c_2}(\boldsymbol{k}+\boldsymbol{q}_{i_1}+\boldsymbol{q}_{i_2})} \cdots
 \nonumber \\
&\qquad
\cdots\times
\Braket{u^{\hat c_{d-1}}(\boldsymbol{k}+\boldsymbol{q}_{i_1}+\cdots +\boldsymbol{q}_{i_{d-1}})|u^{\hat b}(\boldsymbol{k}+\boldsymbol{q}_{i_1}+\cdots +\boldsymbol{q}_{i_{d}})}
\nonumber \\
&=
\sum_{\hat c_1, \dots, \hat c_{d-1} = 1}^{N_-}
(\boldsymbol{q}_1 \wedge \cdots \wedge \boldsymbol{q}_d) \epsilon^{\mu_1 \dots \mu_d}
\frac{\mathcal{F}^{\hat a \hat c_1}_{\mu_1\mu_2}}{2} \cdots \frac{\mathcal{F}^{\hat c_{d-1} \hat b}_{\mu_{d-1}\mu_d}}{2} + O(q^{d+1}).
\label{Conj-even}
\end{align}

As in the (2+1)d case, 
the main result \eqref{main result} 
can further be simplified assuming the uniform distribution
of the Berry curvature and written in terms 
of the bulk Chern number.
Specifically,  we assume 
\begin{align}
    \epsilon^{\mu_1 \dots \mu_d}
  \sum_{\hat c_1, \dots, \hat c_{d/2-1} = 1}^{N_-}
  \mathcal{F}^{\hat a \hat c_1}_{\mu_1\mu_2} \cdots
  \mathcal{F}^{\hat c_{d/2-1} \hat b}_{\mu_{d-1}\mu_d}
  =
    \epsilon^{\mu_1 \dots \mu_d}
  [
  \mathcal{F}_{\mu_1\mu_2} \cdots
  \mathcal{F}_{\mu_{d-1}\mu_d}
  ]^{\hat{a}\hat{b}}
  =
  \delta^{\hat{a}\hat{b}} C
\end{align}
where $C$ is $\boldsymbol{k}$-independent. 
By taking the trace and integral of the both sides, 
we determine the constant as
\begin{align}
  &
    \int \dd^d\boldsymbol{k}\, 
    \epsilon^{\mu_1 \dots \mu_d}
    \mathrm{Tr}_{occ}
    [
    \mathcal{F}_{\mu_1\mu_2} \cdots
    \mathcal{F}_{\mu_{d-1}\mu_d}
    ]
    =
    \frac{(2\pi)^d}{\Omega}
    N_- 
    C
    \nonumber \\
    &\Longrightarrow
    C =
    \rho^{-1}_0
    2^{d/2}
    (d/2)!
    \left(
    \frac{-\ii}{2\pi }
    \right)^{d/2} {\it Ch}_{d/2}, 
\end{align}
where the trace is taken over the 
filled bands,  
$(2\pi)^d/\Omega$ is the volume of the
Brillouin zone,
and 
$\rho_0 = N_-/\Omega$ is
the average density. 
With the assumption,
the RHS of the main result
\eqref{main result}
simplifies as
\begin{align}
&
    \epsilon^{\mu_1 \dots \mu_d}
    \int_{\mathrm{BZ}}
    \frac{\dd^{d}\boldsymbol{k}}{(2\pi)^d}
    \sum_{\hat a, \hat b=1}^{N_-} 
    \sum_{\hat c_1, \dots, \hat c_{d/2-1} = 1}^{N_-}
    \hat{\chi}^{\dag}_{\hat a}(\boldsymbol{k})
    \mathcal{F}^{\hat a \hat c_1}_{\mu_1\mu_2} \cdots
    \mathcal{F}^{\hat c_{d/2-1} \hat b}_{\mu_{d-1}\mu_d}
    \hat{\chi}_{\hat b}(\boldsymbol{k}
    +\boldsymbol{q}_1+\cdots +\boldsymbol{q}_d
    ) 
    \nonumber \\
    &=
    C
    \int_{\mathrm{BZ}}
    \frac{\dd^{d}\boldsymbol{k}}{(2\pi)^d}
    \sum_{\hat a, \hat b=1}^{N_-} 
    \hat{\chi}^{\dag}_{\hat a}(\boldsymbol{k})
    \delta^{\hat a \hat b}
    \hat{\chi}_{\hat b}(\boldsymbol{k}
    +\boldsymbol{q}_1+\cdots +\boldsymbol{q}_d
    ), 
\end{align}
leading to Eq.\ \eqref{main result 2}.

\section{The Wannier functions and the bulk-boundary relation}
\label{The Wannier functions and the bulk-boundary relation}

We have used the Bloch wave functions to define the projected density operator 
and discussed the algebraic structure. 
An alternative approach is to use the Wannier functions.
There, we can avoid the global issue of the Bloch functions. 
The topological nature of bands with non-zero Chern numbers
nevertheless manifests as an obstruction to localizing the Wannier functions.
Another advantage of the Wannier function is 
that it allows us to introduce a boundary to the system
in a natural way.
We can then discuss the boundary density operator algebra.
To illustrate these points, we use the lowest Landau level (LLL) of the (2+1)d QHE.
The following discussion is essentially 
the review of Ref.\ \cite{Azuma_1994}.
We will review the derivation of the 
the GMP algebra (the $W_{1+\infty}$ algebra),
and also 
how
the bulk GMP algebra is related to
the $U(1)$ Kac-Moody algebra realized on the edge of
the integer QHE.
We also discuss the extension to 
the lattice models, i.e., Chern insulators. 

The bulk and boundary algebras are related as follows:
We consider a ground state of integer quantum hall fluid
in the presence of a boundary,
which is obtained by filling a set of Wannier functions. 
The density operators are normal-ordered
with respect to this ground state.
With the normal ordering, the algebra of the density operators acquires a central extension,
which leads to,
in the long wavelength limit,
the Kac-Moody current algebra for (smeared)
density operators near the edge.
It should be noted that
only the long wavelength limit of the $W_{1+\infty}$ algebra, the $w_{\infty}$ algebra,
which is the classical version of the $W_{1+\infty}$ algebra is needed to derive the Kac-Moody algebra at the edge.

We collect, as a start, known relevant results in the QHE.
We will mostly focus on the integer QHE.
The Hamiltonian of a single particle with mass $m$
subjected to a uniform magnetic field $B$ is
$
\mathcal{H} = \frac{1}{2 m}(\boldsymbol{p} + e \boldsymbol{A})^2,
$
where $\partial_x A_y - \partial_y A_x = B$.
We shall adopt the Landau gauge $\boldsymbol{A} = (-By, 0)$,
which is convenient for our later discussion
on a semi-infinite geometry
with a straight edge located at $y=0$.
The LLL wave functions are given by
\begin{align}
\phi_{k_x}(x,y) = \frac{1}{(\pi \ell^2_0)^{\frac{1}{4}}}
 e^{\ii k_x x} e^{-\frac{1}{2\ell^2_0}(y-\ell^2_0 k_x)^2}
\label{LLL}
\end{align}
with normalization $\braket{\phi_{k_x}|\phi_{k_x'}} =
\int_{-\infty}^{\infty} d x \int_{-\infty}^{\infty}\dd{y} \phi^{*}_{k_x}(x,y) \phi_{k_x'}(x,y) = (2 \pi) \delta(k_x-k_x')$.
Here, $\ell_0 = 1/\sqrt{eB}$ is the magnetic length.

Let us assume that
a real space geometry is an infinite cylinder
of circumference $L$,
$(x,y) \in S^1 \times \mathbb{R}$,
with the periodic boundary condition in the $x$-direction, $x=x+L$.
We will eventually take the $L \to \infty$ limit.
The LLL wave function with the wave number $k_x\in \mathbb{R}$
is localized in the $y$-direction
around $\braket{y} = k_x/\ell^2_0$
with its width given by $\braket{(\delta y)^2} \sim \ell^2_0$.

\subsection{The bulk algebra (the GMP algebra)}

We now move on to the second quantization formalism.
The fermion field operators within the LLL are expanded as
\begin{align}
\hat{\psi}(x,y) = \int \frac{\dd{k}_x}{2 \pi} \hat{c}_{k_x} \phi_{k_x}(x,y),
\end{align}
where $\hat{c}_{k_x}$ is the LLL fermion annihilation operator with momentum $k_x$.
The $W_{1+\infty}$ algebra emerges as the algebra of unitary transformations
preserving the LLL condition and the particle number.
Such a unitary transformation is given by
$
\hat{c}_{k_x} \mapsto \hat{c}'_{k_x} = \sum_{k_x} U_{k^{\ }_x,k_x'} 
\hat{c}_{k_x'}.
$
Generators of the unitary transformations are given by
\begin{align}
\hat{\rho}(f) = \int \dd{x} \dd{y}\, \hat{\rho}(x,y) f(x,y),
\end{align}
where $\hat{\rho}(x,y)$ is the density operator projected to the LLL,
$\hat{\rho}(x,y) = \hat{\psi}^{\dag}(x,y) \hat{\psi}(x,y)$,
and $f(x,y)$ is an arbitrary envelope function.
Of our interest is the algebra of the generators $\rho(f)$.
We work with the Fourier components of $\hat{\rho}(x,y)$,
which is given by $\hat{\rho}(f)$ with $f(x,y)= e^{- \ii \boldsymbol{q}\cdot \boldsymbol{x}}$,
\begin{align}
\hat{\rho}(\boldsymbol{q})
&:= \int \dd^2 x\, 
\hat{\rho}(\boldsymbol{x}) e^{-\ii  \boldsymbol{q} \cdot \boldsymbol{x}}
\nonumber \\
&= \int \frac{\dd{k}_x}{2 \pi}\,
\hat{c}^{\dag}_{k_x} \hat{c}^{\ }_{k_x+q_x}\, 
e^{-\frac{\ii \ell^2_0}{2}q_y(2k_x+q_x)} e^{-\frac{\ell^2_0}{4}q^2},
\end{align}
with $\boldsymbol{x}=(x,y)$ and $\boldsymbol{q} = (q_x,q_y)$.

We are interested in the commutator 
of the density operator,
$[\hat{\rho}(\boldsymbol{q}), \hat{\rho}(\boldsymbol{q}')]$. 
We recall that,
for second-quantized operators
$
\hat{Q}_0(A) = \sum_{mn} A_{mn} \hat{\psi}_m^{\dag} \hat{\psi}^{\ }_n,
$
the commutator between $\hat{Q}_0(A)$ and $\hat{Q}_0(B)$ is given by
$[\hat{Q}_0(A), \hat{Q}_0(B)] = \hat{Q}_0([A,B]).$
See Section \ref{n-bracket and second quantization} 
for a more systematic discussion of the formula of 
this type.
For our purpose,
$\hat{Q}_0(A)=\hat{\rho}(\boldsymbol{q})$ and 
$\hat{Q}_0(B)=\hat{\rho}(\boldsymbol{q}')$.
Furthermore, 
introducing a notation for matrix components
of a second quantized operator $\hat O$ as $[\hat O]_{n,m}$
where $\{n,m\}$ represent one-particle basis, 
$A=[\hat{\rho}(\boldsymbol{q})]$
and
$B=[\hat{\rho}(\boldsymbol{q}')]$.
Explicitly, 
they are given by
\begin{align}
[\hat{\rho}(\boldsymbol{q})]_{k_x,k_x'}
= (2 \pi) \delta(k_x'-k_x-q_x) e^{-\frac{\ii\ell^2_0}{2}q_y(k_x+k_x')} e^{-\frac{\ell^2_0}{4}q^2}.
\label{matrix elem density op}
\end{align}
It is convenient to introduce
$\hat{W}(\boldsymbol{q}) =
\hat{\rho}(\boldsymbol{q}) e^{\frac{\ell^2_0}{4}q^2}$,
which obeys the GMP ($W_{1+\infty}$) algebra :
\begin{align}
[\hat{W}(\boldsymbol{q}),\hat{W}(\boldsymbol{q}')]
&= \int \frac{\dd{k}_x}{2 \pi} \int \frac{\dd{k}_x'}{2 \pi} 
\hat{c}^{\dag}_{k_x} \hat{c}^{\ }_{k_x'} \int \frac{dk_x''}{2 \pi} \Bigg[ 
  [\hat{W}(\boldsymbol{q})]_{k_x,k_x''} 
[\hat{W}(\boldsymbol{q}')]_{k_x'',k_x'} -(\boldsymbol{q} \leftrightarrow \boldsymbol{q}') \Bigg]
\nonumber \\
&= -2 \ii  \sin \left( \frac{\ell^2_0}{2}
\boldsymbol{q}\wedge \boldsymbol{q}'
\right) \hat{W}(\boldsymbol{q}+\boldsymbol{q}').
\label{GMP}
\end{align}
In the long-wavelength limit $\ell_0 q \ll 1$,
the GMP algebra reduces to 
\begin{align}
[\hat{\rho}(\boldsymbol{q}),\, \hat{\rho}(\boldsymbol{q}')]
&\simeq -\ii \ell^2_0\, (
\boldsymbol{q}\wedge \boldsymbol{q}')
\, \hat{\rho}(\boldsymbol{q}+\boldsymbol{q}')
\quad  (\ell_0 q \ll 1).
\end{align}
This is the $w_{\infty}$ algebra,
the algebra of classical area-preserving transformations.
For the smeared density operator,
$\hat{\rho}(f) = \int \dd^2 x f(\boldsymbol{x}) \int \frac{\dd^2 q}{(2\pi)^2} 
\hat{\rho}(\boldsymbol{q}) e^{\ii \boldsymbol{q}\cdot \boldsymbol{x}}$,
the $w_{\infty}$ algebra is written as
\begin{align}
\big[ \hat{\rho}(f_1),\, \hat{\rho}(f_2) \big]
&\simeq  \ii \ell^2_0 \hat{\rho} \big( \{ f_1,f_2\}_{\mathrm{PB}} \big)
\quad (\ell_0 q \ll 1),
\end{align}
where $\{f_1,f_2\}_{\mathrm{PB}} =
\partial_x f_1 \partial_y f_2 -
\partial_y f_1 \partial_x f_2$ is
the Poisson bracket.
If we include higher order terms in $\ell_0 q$, the Poisson bracket is replaced by
the Moyal bracket.

\subsection{The boundary algebra (the $U(1)$ Kac-Moody algebra)}

The discussion so far does not refer to the properties of the ground state(s);
the GMP algebra is simply the algebraic relation obeyed by
the density operator projected to the LLL, and is utterly irrelevant to
which ground state is realized in the LLL, which may depend on
the filling fraction, interactions, etc.
We now specialize in the integer QHE and discuss edge excitations and their current algebra
following Azuma's approach
\cite{Azuma_1994}.

Let us introduce a ground state by occupying the LLL states with $\langle y\rangle <0$
whereas states with $\langle y\rangle >0$ are unoccupied:
\begin{align}
\Ket{GS} := \prod_{k_x<0} \hat{c}^{\dag}_{k_x} \ket{0},
\end{align}
Here, $\ket{0}$ the no particle state 
(i.e., annihilated by all $\hat{c}_{k_x}$)
and one should
recall that the LLL states are localized at $y = \ell^2_0 k_x$.
Here, the state $|GS\rangle$ is viewed as the ground state of the integer QHE in
the presence of a boundary (edge) at around $y=0$.
It should be noted that here we consider the single-particle wave functions in the LLL which are energy eigen functions in the absence of the boundary.

For the specified ground state,
the density operator $\hat{\rho}(x,y)$ is not
well-defined due to the ultraviolet (UV) divergence that comes from the Dirac (Fermi) sea.
A well-defined density operator can nevertheless be introduced by
taking the normal order, $\normord{\hat{\rho}(x,y)}$.
Here, the normal order for one-particle operators is defined by
\begin{align}
\normord{ \hat{c}^{\dag}_{k^{\ }_x} \hat{c}^{\ }_{k_x'} } =
\left\{
\begin{array}{ll}
- \hat{c}^{\ }_{k^{\ }_x} \hat{c}^{\dag}_{k_x'}, & (k^{\ }_x=k_x' < 0) , \\
\hat{c}^{\dag}_{k^{\ }_x} \hat{c}^{\ }_{k_x'}, & (\mathrm{otherwise}).
\end{array} \right.
\end{align}

Let us now consider 
the commutator of 
the normal-ordered density operator 
$\normord{\hat{\rho}(q_x,y)}$,
$[\normord{\hat{\rho}(q_x,y)},\normord{\hat{\rho}(q'_x,y')}]$. 
Here,
$\hat{\rho}(q_x,y)$ is
the Fourier transformation 
of
$\hat{\rho}(x,y)$
along the $x$-direction and given by 
\begin{align}
\hat{\rho}(q_x,y)
&= \int \frac{\dd{k}_x}{2 \pi} 
\hat{c}^{\dag}_{k_x} \braket{\tilde \phi_{k_x}|y} \braket{y|\tilde \phi_{k_x+q_x}} 
\hat{c}_{k_x+q_x},
\end{align}
where we introduced the basis of the one-particle states,
\begin{align}
\braket{y|\tilde \phi_{k_x}} = \frac{1}{(\pi \ell^2_0)^{\frac{1}{4}}} 
e^{-\frac{1}{2 \ell^2_0} (y- \ell^2_0 k_x)^2}.
\end{align}
Note that 
$\braket{\tilde \phi_{k_x} | y} \braket{y | \tilde \phi_{k_x+q_x}}$,
and 
hence
$\hat{\rho}(q_x,y)$,
is exponentially localized around $y \sim \ell^2_0(k_x+\frac{q_x}{2})$ as
\begin{align}
\braket{\tilde \phi_{k_x}|y} \braket{y|\tilde \phi_{k_x+q_x}}
= \frac{1}{(\pi \ell^2_0)^{\frac{1}{2}}} 
e^{-\frac{1}{\ell^2_0}(y-\ell^2_0(k_x+\frac{q_x}{2}))^2} 
e^{-\frac{\ell^2_0}{4} q_x^2}.
\end{align}

We now recall, for two normal-ordered operators 
$\hat{Q}(A)=\normord{\hat{Q}_0(A)}$ and 
$\hat{Q}(B)=\normord{\hat{Q}_0(B)}$,
their commutator is given by
$
[\hat{Q}(A), \hat{Q}(B)] = \hat{Q}([A,B]) + S_2(A,B),
$
where
we consider a many-body ground state in the Fock space
given by
$\ket{GS} = \prod_{n<0} \hat{\psi}^{\dag}_n \ket{0}$, 
and $S_2(A,B)$ is a $c$-number and given by 
\begin{align}
S_2(A,B) = \sum_{n<0, m>0} \Big( A_{nm} B_{mn} - B_{nm} A_{mn} \Big).
\label{S2 1}
\end{align}
The systematic discussion of the formula of type Eq.\ \eqref{S2 1}
is given in Section 
\ref{n-bracket and second quantization}. 
Specializing to our case, 
the commutator 
$[\normord{\hat{\rho}(q_x,y)},\normord{\hat{\rho}(q'_x,y')}]$ 
is given by 
\begin{align}
[\normord{\hat{\rho}(q_x,y)},\normord{\hat{\rho}(q'_x,y')}]
  &= \hat{Q}([[\hat{\rho}(q_x,y)], [\hat{\rho}(q_x',y')]])
    + S_2([\hat{\rho}(q_x,y)], [\hat{\rho}(q_x',y')]), 
\label{s2} 
\end{align}
where 
the matrix elements of $\hat{\rho}(q_x,y)$ are given by 
\begin{align}
[\hat{\rho}(q_x,y)]_{k_x,k_x'} = (2 \pi) \delta(k_x'-k_x-q_x) \braket{\tilde \phi_{k_x}|y} \braket{y|\tilde \phi_{k_x+q_x}}. 
\end{align}

Following Ref.\ \cite{Martinez_Stone_1993},
we will further consider to take a spatial average of $\normord{\hat{\rho}(q_x,y)}$,
and introduce 
\begin{align}
\hat{\varrho}(q_x)
&:= \int_{- L}^{L} \dd{y}\, \normord{\hat{\rho}(q_x,y)}
\nonumber \\
&= \int^{\Lambda}_{-\Lambda} \frac{\dd{k}_x}{2 \pi} 
\normord{\hat{c}^{\dag}_{k_x} \hat{c}_{k_x+q_x}} \braket{\tilde \phi_{k_x}|P_{L}| \tilde \phi_{k_x+q_x}}
\end{align}
with $P_{L} = \int_{-L}^{L}\dd{y} \ket{y} \bra{y}$ and $L \gg \ell_0$.
The operator $\hat{\varrho}(q_x)$ is interpreted as 
the density operator which creates excitations at the boundary
$y=0$. 
We will eventually send $\Lambda\to \infty$ and $L\to \infty$. 
Note that in this limit, 
$
\int \dd{y} \hat{\rho}(q_x,y)
= \int \frac{\dd{k}_x}{2 \pi} 
\hat{c}^{\dag}_{k_x} \hat{c}_{k_x+q_x} e^{-\frac{\ell^2_0}{4} q_x^2}
= \int \frac{\dd{k}_x}{2 \pi} 
\hat{c}^{\dag}_{k_x} \hat{c}_{k_x+q_x} + O(q_x^2).
$

Let us now evaluate the first term in the RHS of Eq.\ \eqref{s2}.  
The relevant matrix elements of the first term are 
\begin{align}
&
\big[[\hat{\rho}(q_x,y)], [\hat{\rho}(q_x',y')]\big]_{k_x,k_x'}
\nonumber \\
&= (2 \pi) \delta(k_x'-k_x-q_x-q_x') \braket{\tilde \phi_{k_x} | y} \braket{y | \tilde \phi_{k_x+q_x}} \braket{\tilde \phi_{k_x+q_x} | y'} \braket{y' | \tilde \phi_{k_x+q_x+q_x'}}
- (q_x \leftrightarrow q_x').
\end{align}
Recall that $\braket{\tilde \phi_{k_x} | y} \braket{y | \tilde \phi_{k_x+q_x}}$ 
is exponentially localized around $y \sim \ell^2_0(k_x+\frac{q_x}{2})$.
Upon taking the spatial average in the $y$-direction,
the matrix elements of $[[\hat{\varrho}(q_x)], [\hat{\varrho}(q_x')]]$ 
may have contributions near $k_x, k_x' \sim L/\ell^2_0$.
We can take the limit $L \to \infty$ for an arbitrary fixed $k_x \in \mathbb{R}$, then
$
\big[[\hat{\varrho}(q_x)], [\hat{\varrho}(q_x')]\big]_{k_x,k_x'} \to 0 \ \ (L \to \infty).
$
Thus, the first term in the RHS of Eq.\
\eqref{s2}, upon taking the spatial average in the $y$-direction,
vanishes. 
As for the second term (the $c$-number part) 
of
Eq.\ \eqref{s2}, 
it 
can be computed as
\begin{align}
S_2([\hat{\rho}(q_x,y)], [\hat{\rho}(q_x',y')])
= \delta(q_x+q_x') \int_{-q_x}^{0} \dd{k}_x \braket{\tilde \phi_{k_x} | y} \braket{y | \tilde \phi_{k_x+q_x}} \braket{\tilde \phi_{k_x+q_x} | y'} \braket{y' | \tilde \phi_{k_x}} .
\end{align}
Upon taking the spatial average in the $y$-direction,
for $\ell^2_0 q_x \ll L$, we have
\begin{align}
S_2([\hat{\varrho}(q_x)], [\hat{\varrho}(q_x')])
&= \delta(q_x+q_x') \int_{-q_x}^{0} \dd{k}_x | \braket{\tilde \phi_{k_x} | P_{L} | \tilde \phi_{k_x+q_x}} |^2
\nonumber \\
&\to \delta(q_x+q_x') \int_{-q_x}^{0} \dd{k}_x | \braket{\tilde \phi_{k_x} | \tilde \phi_{k_x+q_x}} |^2 \ \ (L \to \infty)
\nonumber \\
&= \delta(q_x+q_x') q_x e^{-\frac{\ell^2_0}{2} q_x^2}
\nonumber \\
&= \delta(q_x+q_x') q_x (1 + O( (\ell_0 q_x)^2 )).
\end{align}
Note that the $\delta(q_x + q_x') q_x$ factor arises from the condition of momenta
$k_x' = k_x + q_x > 0$
and 
$
k_x = k_x +q_x' < 0$. 
The microscopic structure of wave function $\ket{\tilde \phi_{k_x}}$ is not relevant.

Summarizing, 
the smeared density operator $\varrho(q_x)$ obeys 
the algebra
\begin{align}
\big[\varrho(q_x),\, \varrho(q_x')\big] = \delta(q_x + q_x') q_x + O( (\ell_0 q_x)^2 ), 
\end{align}
at low energies. This is nothing but the $U(1)$ Kac-Moody algebra.

\subsection{Beyond the (2+1)d Landau levels}

\subsubsection{(2+1)d Chern insulators}
\label{2+1d Chern insulators}

The above calculation for the (2+1)d QHE in the Landau gauge can be generalized
to lattice systems (Chern insulators) and to higher dimensions by using the
hybrid Wannier functions
\cite{Qi2011}.
Let us start with the tight-binding model \eqref{tight binding}.
For our purpose of introducing a straight boundary along the $x$-direction at around $y\sim 0$, it is convenient to consider the geometry which is periodic in the $x$-direction, and infinitely long in the $y$-direction. 
I.e., the geometry is an infinite cylinder. 
For simplicity, we consider only a single band $\ket{u(k_x,k_y)}$ to be occupied (and hence we drop the band index), and we set lattice constants as $a_x = a_y = 1$.
The Bloch wave function can be chosen to obey a smooth and periodic gauge for the $k_y$-direction $\ket{u(k_x,k_y+2\pi)}=\ket{u(k_x,k_y)}$. 
With this gauge choice, the hybrid Wannier function 
(which is exponentially localized along the $y$-direction) is defined by 
\begin{align}
    \ket{w(k_x,n_y)} := \oint \frac{\dd{k}_y}{2 \pi} e^{\ii k_y (\hat n_y - n_y)} \ket{u(k_x,k_y)}, 
\end{align}
where $\hat n_y$ is the $y$-component of the coordinate operator.
The mean-localized position of $\ket{w(k_x,n_y)}$ is given by 
\begin{align}
\Braket{w(k_x,n_y)| \hat n_y | w(k_x,n_y)} = n_y + \frac{\theta_y(k_x)}{2 \pi},
\end{align}
where $\theta_y(k_x)$ is the $k_x$-resolved polarization along the $y$-direction 
\begin{align}
\theta_y(k_x) = \ii \oint_0^{2\pi} \dd{k}_y \braket{u(k_x,k_y) | \partial_{k_y} u(k_x,k_y)}. 
\end{align}
With this, the 1st Chern number is given by $Ch_1 = \frac{1}{2\pi} \oint_0^{2\pi} d_{k_x} \theta_y(k_x) \in \Z$. 

Under the large gauge transformation of the Bloch wave functions $\ket{u(k_x,k_y)} \mapsto \ket{u(k_x,k_y)} e^{-\ii n_y k_y}$, the polarization shifts as
\begin{align}
\frac{\theta(k_x)}{2 \pi} \mapsto \frac{\theta(k_x)}{2 \pi} + n_y.
\end{align}
Thus, we can always choose the gauge of $\ket{u(k_x,k_y)}$ satisfying $\theta(0) \in [-\pi,\pi]$.
It is crucial to observe that since $\ket{w(k_x,n_y)}$ satisfies
\begin{align}
\ket{w(k_x + 2 \pi,n_y)} = \ket{w(k_x,n_y+Ch_1)},
\end{align}
the $k_x$-resolved polarization satisfies 
\begin{align}
\frac{\theta(k_x + 2 \pi)}{2 \pi} - \frac{\theta(k_x)}{2 \pi} = Ch_1.
\end{align}
This means that, by changing $k_x$ continuously (adiabatically) for
$-\infty < k_x <+\infty$, one can span entirely the space of all occupied single-particle states. 
When $Ch_1=1$, this is essentially the behavior of the Landau level wavefunctions (\ref{LLL}) in the Landau gauge. 
Note that $k_x$ is originally defined in the BZ, but we have extended 
it to value in $\mathbb{R}\simeq [-\infty, +\infty]$. 
For simplicity, 
let us consider the case with $Ch_1 = 1$.
The one-dimensional Wannier states $\ket{w(k_x,n_y)}$
can be labeled by single $k_x \in \mathbb{R}$ 
because of the relation $\ket{w(k_x+2 \pi,n_y)} = \ket{w(k_x,n_y+1)}$.
We thus define
$
\ket{W_{k_x}} := \ket{w(k_x,n_y=0)} \ \ (k_x \in \mathbb{R}).
\label{Wannier}
$
The localized position of $\ket{W_{k_x}}$ is $\braket{W_{k_x} | \hat n_y | W_{k_x}} = \frac{\theta(k_x)}{2 \pi} \in \mathbb{R}$.
Using $\ket{W_{k_x}}$, we can essentially repeat the calculations in 
the previous section, i.e.,  we can define the 
projected density operator, the ground state in the presence of a 
boundary by filling with $\ket{W_{k_x}}$ with $k_x<0$, 
and calculate the algebra of the projected and normal-ordered
density operator.
The Wannier functions can also be used to discuss 
higher-dimensional topological insulators on a lattice, 
and in the continuum. 

\subsubsection{The (4+1)-dimensional lowest Landau level}

Let us have a closer look 
a particular continuum model realizing a (4+1)-dimensional topological insulator.
A convenient model for us is the (4+1)-dimensional analog of the
Landau levels
introduced in Ref.\ \cite{2013PhRvL.111r6803L},
which is very similar to the (2+1)-dimensional Landau levels.
One caveat of this model is that it is not translation invariant
in all directions (and hence may be inconvenient to discuss
the bulk density operator algebra).
On the other hand, for the purpose of discussing
the bulk-boundary relation, the lack of translation invariance in one direction is not a problem.
 
The model of the (4+1)-dimensional Landau levels is given by
\begin{align}
  \mathcal{H}
  =
  \frac{-\partial^2_4}{2m}
  +
  \frac{m\omega^2}{2}
  \left[
  x_4
  - \ell^2_0
  \Big(-\ii \sum_{i=1}^3 \partial_i \sigma_i\Big)
  \right]^2
  \label{4d LLL model}
\end{align}
with
$\omega = 1/(m\ell_0)$.
The energy eigenstates are labeled by
the integer $n=0,1,2,\ldots$ (the "Landau level index"),
the three-dimensional momentum
$\mathbf{k}=(k_1,k_2,k_3)$,
and the helicity $a=\pm$.
The energy eigenvalues are given by
$\epsilon^{na}(\mathbf{k})
=
(n+1/2) \omega$.
I.e., for each Landau level, the spectrum is "flat"
independent of $\mathbf{k}$ and $a$.
Placing the chemical potential in between 
Landau levels to realize a band insulator,
this model
exhibits the quantized non-linear electromagnetic response,
which is the four-dimensional generalization of the QHE.
If we impose
open boundary conditions in the $x_4$-direction,
the system supports
Weyl fermions with opposite chiralities
on the two three-dimensional boundaries, respectively.

In the following, we shall focus on the LLL, $n=0$.
The LLL wave functions are given by
\begin{align}
 \vec{\phi}_{\mathbf{k},a} (\mathbf{x}, x_4)
 =
 \tilde{\phi}_{\mathbf{k},a} (x_4)
 \cdot 
 e^{ \ii \mathbf{k}\cdot \mathbf{x}} \vec{e}_a (\mathbf{k})
 =
 \frac{1}{(\pi \ell^2_0)^{\frac{1}{4}}}
 e^{ - \frac{1}{2\ell^2_0}
 ( x_4 - \ell^2_0 \lambda_a(\mathbf{k}) )^2 }
 e^{ \ii \mathbf{k}\cdot \mathbf{x}} \vec{e}_a (\mathbf{k}),
\end{align}
where the two-component spinor $\vec{e}_a(\mathbf{k})$ is an eigenstate of $\mathbf{k}\cdot \boldsymbol{\sigma}$,
$(\mathbf{k}\cdot \boldsymbol{\sigma}) \vec{e}_a (\mathbf{k})
  =
  \lambda_a (\mathbf{k}) \vec{e}_a (\mathbf{k})$,
$\lambda_{\pm}(\mathbf{k}) = \pm |\mathbf{k}|$.
The LLL wave functions are localized in the $x_4$-direction at around
$
  \langle \phi_{\mathbf{k},a} | \hat{x}_4 | \phi_{\mathbf{k},a} \rangle
  =
  \ell^2_0 \lambda_a (\mathbf{k}). 
$
The electron density operator projected onto the LLL 
and its Fourier transform 
is given by
\begin{align}
  \hat{\rho}(\mathbf{q})
  &=
  \int \dd{x}_4\,
  \hat{\rho}(\mathbf{q}, x_4)
  =
 \int \dd{x}_4\, \dd^3{\bf x}\,
  e^{- \ii {\bf q}\cdot {\bf x}}
  \hat{\rho}({\bf x}, x_4)
  \nonumber\\
  &=
  \sum_{\mathbf{k},a,b}
  \hat{c}^{\dag}_{\mathbf{k}a}
  \,
  e^{-\frac{\ell^2_0}{4} 
  (\lambda_b(\mathbf{k}+\mathbf{q}) - \lambda_a(\mathbf{k}) )^2 } 
  \vec{e}_{a}(\mathbf{k})^*\cdot \vec{e}_b(\mathbf{k}+\mathbf{q})
  \,
  \hat{c}^{\ }_{\mathbf{k}+\mathbf{q}b},
\end{align}
where $\hat{c}^{\dag}_{\mathbf{k} a}$ is the 
electron creation operator for the LLL, 
and we have averaged over the $x_4$ direction.
In the long-wavelength limit
$\mathbf{q}\to\mathbf{0}$,
the projected density operator is given by
$\hat{\rho}(\mathbf{q}) \sim 
\sum_{\mathbf{k},a} \hat{c}^{\dag}_{\mathbf{k},a} \hat{c}^{\ }_{\mathbf{k}+\mathbf{q},a}$.

Following Azuma's approach, let us now consider a ground state by filling the LLL states with
$\langle \hat{x}_4 \rangle < 0$,
$
  |GS \rangle =
  \prod_{\mathbf{k}} 
  \hat{c}^{\dag}_{\mathbf{k}-}
  |0\rangle,
$
with the grading operator
$F = 2 P_+ -1 = 2(1-P_-) -1 = 1- 2P_-$
given by
\begin{align}
  F  =
  \frac{ \sum_{i=1}^3 k_i \sigma_i}{|\mathbf{k}|}
  =
  \mathrm{sign}\, \mathcal{H}_{{\it Weyl}}
  \label{grading op}.
\end{align}
Now, to consider the algebra of the normal-ordered, projected density operator
we need to evaluate $S_4$ in Eq.\ \eqref{S4}
using the grading operator \eqref{grading op}.
This problem is identical to evaluating the density operator algebra within the (3+1)-dimensional Weyl fermion with the single-particle Hamiltonian 
$
\mathcal{H}_{Weyl} = -\ii \sum_{i=1}^3 \sigma_i \partial_i
$.
While we have treated the specific model of the Landau levels,
Eq.\ \eqref{4d LLL model},
similar bulk-boundary correspondence can be established for more generic models.

\section{Discussion}
\label{discussion}

In this paper, we identified the higher-bracket structure for the algebra of 
the projected density operators in topological insulators, both for their bulk and boundaries. 
For the density operator on the boundary of topological insulators, 
we identified Schwinger-term-like c-number parts (e.g., $S_4$), which are topological invariants;
they are the so-called cyclic cocycles in noncommutative geometry of Connes. 
At the non-interacting level, they play central roles in the noncommutative geometry approach 
to address the effects of disorder
\cite{belissard_JMatPhys94, Hastings_annals_physics11, loring_hastings_EPL_10, Prodan_2013}.
Our work finds a many-body (an interacting) counterpart of  
the noncommutative geometry approach to topological insulators.

There are many further issues to be discussed. 
Below, we list some of the significant further problems. 

-- 
Guided by the mathematical consistency 
(i.e., the generalized Jacobi identities)
we proposed to drop $R_4$ (and its higher dimensional counterparts) by a suitable renormalization.
It is nevertheless desirable to have a better 
physical understanding of the renormalization.
We also note that it would also be possible to guide ourselves by using alternative 
mathematical structures than higher Lie algebras,
such as $n$-Lie algebras 
\cite{filippov_n-lie_1985},
in which 
a different definition of the generalized Jacobi identity
is used.
See, for example, Ref.\ \cite{DeBellis_2010} and references there.

--
In this paper, we focused on $d={\it even}$ bulk spatial dimensions.
It is interesting to study the case of $d={\it odd}$ bulk spatial dimensions.
Topological insulators in these dimensions need to be protected by 
some symmetry, such as chiral symmetry (i.e., symmetry class AIII).
We note that this is a statement at the level of topological band theory --
with interactions, the topological classification may be 
altered (reduced) or can be completely different in principle.
With these remarks in mind, we can still look 
at the algebraic structure of (projected/normal-ordered)
density operators. With chiral symmetry, 
relevant electron coordinates are 
the chiral coordinate operators \cite{Shiozaki_2013},
Similarly, we need to consider the chiral density operators
as relevant second-quantized operators. 
The bulk and boundary algebras 
for the chiral density operators
can be calculated in 
a way similar to the calculations presented in this paper
\cite{LangmannUnpublished}.

--
Looking ahead, despite all these issues, we believe the higher bracket structure of
the density operator would play an important role 
in the description of topological insulators -- in particular in the presence 
of interactions. 
Here, it is worth recalling the crucial role played by the 
GMP algebra and the $U(1)$ current algebra in the context of 
the integer as well as fractional quantum Hall effects.
The GMP algebra is relevant in describing bulk excitations such as magneto roton in the QHE;
The current algebra of edge excitations that arise 
from the bulk GMP algebra upon inclusion of its central extension 
can be used as 
a spectrum-generating algebra.
In a sense, the GMP algebra provides a collective or "hydrodynamic"
description of quantum Hall states. In a similar vein, a natural next step following the present paper is
an application of the $n$-bracket structure of the density operators to describe the bulk and boundary
excitations of topological insulators, 
and their interacting counterparts, putative fractional topological
insulators.
It is therefore important to study 
if the $n$-bracket of the density operators,
computed in this paper in the limit of vanishing interactions,
can receive some renormalization or interaction corrections.
Even in the absence of interactions, one may wonder to which extent the results of the present paper
for the $n$-bracket of the density operators is robust,
e.g., in the presence of disorder, or beyond the semiclassical treatment adopted in this paper.
Possible connections with quantum anomalies should also be explored.

-- 
One of the most pressing issues is the application of
the higher-bracket structure of the density operators. 
Are there any physical observables or phenomena associated with the higher algebraic structure?
To help find applications,
in a separate paper
\cite{LangmannUnpublished}, 
we plan to visit the algebra of (boundary) current operators.
There, we consider the smeared 
current operators
\begin{align}
\hat{J}_{\mu}(f)  =
\int_{\partial M_4} \dd^3\mathbf{x}\,
f_{ab}(\mathbf{x})
\normord{
\hat{\psi}^{\dag}_a(\mathbf{x})
\sigma_{\mu}
\hat{\psi}^{\ }_{b}(\mathbf{x})},
\quad \mu=0,\ldots,3,
\end{align}
(We here introduce 
"color" or "flavor" indices $a,b=1,\cdots, N$, and 
consider non-Abelian currents.) 
We will show that
the same c-number part $S_4$,
appeared in the $4$-bracket of 
the density operator,
also appears in the repeated commutator of the current operators, 
\begin{align}
&
\frac{1}{8}
\epsilon^{i_0i_1i_2i_3}
\big[
\big[
\big[
\hat{J}_0(f_{i_0} ),
\hat{J}_1(f_{i_1} )
\big],
\hat{J}_2(f_{i_2} )
\big],
\hat{J}_3(f_{i_3})
\big]
\nonumber \\ 
&
\quad
=
{\ii} \hat{J}_0([ f_0, f_1, f_2, f_3] ) +
b_4([f_0, f_1, f_2, f_3]),
\label{current alg in 4D}
\end{align}
where $b_4$ is given by Eq.\ \eqref{def b4}.
As we did for the higher bracket of 
the normal-ordered density operator,
$b_4$ can be split into $S_4$ and $R_4$,
and, by subtracting $R_4$, we can introduce 
a regularized version of the repeated commutator.
While the regular commutator appears naturally 
in the linear response theory, 
the repeated commutator structure may 
play a role in 
the non-linear response of the system:
The change in an observable 
induced by a perturbation $\hat{V}$
is
$
\delta\langle \hat{O}(t) \rangle
=
\langle \hat{O}(t) \rangle
-
\langle \hat{O}(0) \rangle
=
+\ii \int^t dt'
\langle
[
\hat{V}(t'),\hat{O}]
\rangle
+
( \ii^2/2!) 
\int^t dt' \int^t dt^{\prime\prime}
\langle
[\hat{V}(t^{\prime\prime}),[\hat{V}(t'), \hat{O}]]
\rangle
+
\cdots
$.

--
Finally, 
let us discuss a possible field theory description 
of the higher-bracket structure. 
We recall that,
while the geomertrical aspects such as 
the GMP algebra are not captured by 
topological field theories,
one can formulate the non-commutative Chern-Simons theory
\cite{susskind2001quantum}
that can capture both topological and geometrical aspects of  
integer as well as fractional quantum Hall states
\cite{Lapa_2018}.
Here, we
use the functional bosonization and derive 
the (hydrodynamic) effective field theory of
$(d+1)$d topological insulators
\cite{Chan_2013}.
In this approach, the generating functional of 
correlation functions of $U(1)$ currents,
obtained by the path integral over 
the fermion (electron) field $\psi^{\dag},\psi$
in the presence of 
the background $U(1)$ gauge field $A^{\mathrm{ext}}$,
\begin{align}
Z[A^{{\rm ext}}]
=
\int {\cal D}[\psi^{\dag}]{\cal D}[\psi]
\exp \Big [ \ii K_F(\psi^{\dag},\psi, A^{\mathrm{ext}})\Big],
\end{align}
is expressed in terms of the bosonic path integral, 
\begin{align}
Z[A^{{\rm ext}}]
=
\int {\cal }D[a]{\cal D}[b]
Z[a]
\exp \left[\frac{\ii}{2\pi} \int b (da -dA^{\mathrm{ext}})
\right].
\label{bf bz}
\end{align}
Here, 
$K_F[\psi^{\dag},\psi, A^{\mathrm{ext}}]$ 
is the fermionic action describing the (topological) insulator
in the presence of the background $U(1)$ gauge field,
and $a$ and $b$ are dynamical 
1- and ($d-1$)-form gauge fields, respectively.
From the coupling to the background field, 
we read off the bosonization dictionary
in which the $U(1)$ current $j$ is represented 
in term of $b$ as $j\propto db$. 
We also notice the $BF$-type coupling 
between $a$ and $b$ at level 1. 
Let us first discuss the case of 
the integer QHE and Chern insulators in $(2+1)$ dimensions. 
For a band insulator with the first Chern number 
${\it Ch}_1=1$, 
the leading part in 
$Z[a]$ in Eq.\ \eqref{bf bz} is given by the (2+1)d Chern-Simons term, 
$
\ln Z[a]= (i/4 \pi) \int ada
+\cdots.
$
The resulting topological field theory 
(the level-1 ${\it BF}$ theory with the Chern-Simons term $ada$),
$Z=\int {\cal D}[a]{\cal D}[b] \exp i S$
with $S = \int (1/2\pi) b(da-dA^{\mathrm{ext}}) 
+ (1/4\pi) ada$,
correctly reproduces the topological properties, 
but not the geometric ones, such as the GMP algebra.  
One can however include 
higher order terms that appear in $Z[a]$
\cite{Fradkin_2002}, leading to 
\begin{align}
S=\int d^3 x\left[\frac{1}{2\pi} \varepsilon^{\mu \nu \lambda} 
b_\mu \partial_\nu\left(a-A^{ \mathrm{ext} }\right)_\lambda+
\frac{1}{4\pi}
\left(
\varepsilon^{\mu \nu \lambda} a_\mu \partial_\nu a_\lambda+
\frac{\theta}{3}
\{a_{\mu}, a_{\nu}\} a_{\lambda}
\right)
\right]+\cdots
\end{align}
where $\{\cdots \}$ 
is the Poisson bracket, 
$
\{a_{\mu}, a_{\nu}\}=
\varepsilon_{i j} \partial_i a_\mu \partial_j a_\nu$,
and $\theta$ is a dimensionful parameter, 
the noncommutative parameter, inversely proportional to 
the applied magnetic field. 
(Here, $\cdots$ includes even higher order terms and, also, the Maxwell term.)
One recognizes the term 
$
(\theta/3)
\{a_{\mu}, a_{\nu}\} a_{\lambda}
$
as arising from the leading order expansion of the non-commutative Moyal product;
the second part in $S$ agrees with 
the leading order expansion of the non-commutative 
Chern-Simons term, 
$
(1/4\pi) \varepsilon^{\mu\nu\lambda} 
\left(
a_{\mu}\star \partial_{\nu} a_{\lambda}
+
(2i/3) a_{\mu} \star a_{\nu} \star a_{\lambda}
\right)
$.

Turning now to a (4+1)d topological insulator with unit second Chern number ${\it Ch}_2=1$, say,
the topological field theory is given 
in terms of the one-form $a$ and three-form gauge $b$ gauge fields.
The leading topological term of the action is given by the (4+1)d Chern-Simons term, 
\begin{align}
\ln Z[a]  =\frac{\ii}{24 \pi^2} \int d^5 x \epsilon^{\mu \nu \lambda \rho \sigma} a_\mu \partial_\nu a_\lambda \partial_\rho a_\sigma 
+\cdots.
\end{align}
As before, the resulting topological field theory (the level-1 ${\it BF}$ theory with 5d Chern-Simons term) 
only reproduces the topological properties,
and hence one needs to go beyond the leading order Chern-Simons term.  
We have not carried out this calculation.
Nevertheless, a natural guess 
is to replace the Poisson bracket
appearing in the (2+1)d case 
by the Nambu bracket. 
This results in the action
\begin{align} 
S=  \int d^5 x\left[ \varepsilon^{\mu \nu \lambda \rho \sigma} b_{\mu \nu \lambda} \partial_\rho\left(a-A^{\mathrm{ext}}\right)_\sigma
+
\frac{1}{24\pi^2}
\left(
\varepsilon^{\mu \nu \lambda \rho \sigma} a_\mu \partial_\nu a_\lambda \partial_\rho a_\sigma
+\theta \varepsilon^{\mu \nu \lambda \rho \sigma} \varepsilon_{i j k l} \partial_i a_\mu \partial_j a_\nu \partial_k a_\lambda \partial_l a_\rho a_\sigma
\right)\right]+\cdots.
\end{align}
This is our proposal for the field theory description of the higher-bracket structure 
of the density operators in (4+1)d topological insulators.  
The derivation and analysis of this theory are left for future work.

\begin{acknowledgments}
We are grateful to Alan Carey for a remark about higher Lie algebras that prompted us to return to this project and helped us to overcome a deadlock preventing us from finishing this project in 2015. 
S.R. thanks Yichen Hu,
Charlie Kane, Kohei Kawabata, and Pok Man Tam
for discussion on 
the non-linear response and density correlation functions
of a Fermi gas (Fermi liquid)
\cite{Tam_2022,tam2023topological},
which have some similarities
with the density operator algebra in this work.
We thank Alan Carey, Jouko Mickelsson,
Titus Neupert, Mike Stone, Raimar Wulkenhaar, and Peng Ye for useful discussions. 
This work is supported by INSPIRE grant at the University of Illinois at Urbana-Champaign.
K.S. was supported by JSPS Postdoctoral Fellowship for Research Abroad.
S.R.~is supported by the National Science Foundation under  Award No.\ DMR-2001181, and by a Simons Investigator Grant from the Simons Foundation (Award No.~566116).
This work is supported by the Gordon and Betty Moore Foundation through Grant GBMF8685 toward the Princeton theory program. 
\end{acknowledgments}

\appendix 

\bibliography{reference}

\end{document}